\documentclass[twocolumn]{article}

\usepackage{tabularx}
\usepackage{natbib}

\usepackage{graphicx}
\usepackage{dcolumn}
\usepackage{bm}

\usepackage[colorlinks=true,
linkcolor=blue, 
anchorcolor=black, 
citecolor=blue, 
urlcolor=blue
]{hyperref}

\usepackage[a4paper]{geometry}
\usepackage{caption}
\usepackage{authblk}
\usepackage{graphicx}
\usepackage{graphics}
\usepackage{multirow}
\usepackage{longtable}
\usepackage{supertabular}
\usepackage{lscape}
\usepackage{rotating} 
\usepackage{enumitem}
\usepackage{authblk}

\providecommand{\keywords}[1]{\textbf{\textit{Key words---}} #1}
\begin{document}

\title{Behavioral analysis in immersive learning environments: A systematic literature review and research agenda}

\author[1]{Yu Liu}
\author[2]{Kang Yue}
\author[1]{Yue Liu\thanks{Corresponding author: Yue Liu, liuyue@bit.edu.cn.}}

\affil[1]{Beijing Engineering Research Center of Mixed Reality and Advanced Display, School of Optics and Photonics, Beijing Institute of Technology, Beijing 100081, China.}
\affil[2]{Institute of Software, Chinese Academy of Sciences, Beijing 100045, China.}

\date{} 

\maketitle

\begin{abstract}
	The rapid growth of immersive technologies in educational areas has increased research interest in analyzing the specific behavioral patterns of learners in immersive learning environments. Considering the fact that research on the technical affordances of immersive technologies and the pedagogical affordances of behavioral analysis remains fragmented, this study first contributes by developing a conceptual framework that amalgamates learning requirements, specification, evaluation, and iteration into an integrated model to identify learning benefits and potential hurdles of behavioral analysis in immersive learning environments. Then, a systematic review was conducted underpinning the proposed conceptual framework to retrieve valuable empirical evidence from the 40 eligible articles during the last decade. The review findings suggest that (1)  there is an essential need to sufficiently prepare the salient pedagogical requirements to define the specific learning stage, envisage intended cognitive objectives, and specify an appropriate set of learning activities, when developing comprehensive plans on behavioral analysis in immersive learning environments.	(2) Researchers could customize the unique immersive experimental implementation by considering factors from four dimensions: learner, pedagogy, context, and representation. (3) The behavioral patterns constructed in immersive learning environments vary by considering the influence of behavioral analysis techniques, research themes, and immersive technical features. 	(4) The use of behavioral analysis in immersive learning environments faces several challenges from technical, implementation, and data processing perspectives. This study also articulates critical research agenda that could drive future investigation on behavioral analysis in immersive learning environments.
\end{abstract}


\section{Introduction}

Immersion refers to the experience when people become deeply absorbed by such attractive objects as music, movies, works of art, landscapes, or even their own thoughts, and forget their surroundings temporarily \citep{Yuen2013}.
While, in the educational context, especially in the context of e-learning \citep{MONAHAN20081339}, mobile learning \citep{Sun2016}, or ubiquitous learning \citep{Liu2009}, the immersive learning experience has a more specific meaning, i.e., people engage in the mediated or simulated learning environment created by immersive technologies, including augmented reality (AR), virtual reality (VR), mixed reality (MR), and involve the willing suspension of disbelief \citep{Dede2017}.
Immersive technology makes it possible to construct a learning environment that blurs the distinct boundary between the tangible and digital or simulated world, and brings abundant possible advantages in the educational areas that have been extensively studied in numerous previous research, such as enhancing learning experience and emotions \citep{Huang2016}, promotion of learning motivation and creativity \citep{Wei2015}, and experiencing the presence through immersive virtual field trips \citep{Han2021}.

Disambiguation of fundamental terms of immersive technology is needed to describe clear delineation and specific features among the three terms of VR, AR, and MR. 
Augmented reality refers to the technology that enriches the sensorial perception of a person with the superimposition of virtual content anchored in the real world \citep{Daponte2014,Sereno2020,Alkhabra2023}. Azuma has identified three characteristics of AR: the combination of real and virtual content, real-time interactivity, and 3D registration \citep{Azuma1997}. Several AR tracking approaches, such as marker-based, sensor-based, and marker-less tracking methods, have been developed to achieve stable and flexible tracking and registration performance \citep{Wang2016}. 
VR refers to the technology that creates an interactive three-dimensional virtual world that not only produces a faithful reproduction of “reality” but expands the bounds of reality to accomplish things that cannot come true in physical reality \citep{Kardong-Edgren2019,Slater2016}. Based on the criteria proposed by \cite{Slater1997}, the classification of VR can be described into three general categories according to the level of immersion: non-immersive VR, semi-immersive VR, and fully-immersive VR \citep{Rose2018}. 
According to Milgram and Kishino’s reality–virtuality continuum \citep{Milgram1994} and Benford’s taxonomy \citep{Benford1998}, MR refers to the technology that integrates virtual and real worlds as a whole space that spans the local and remote as well as the physical and synthetic dimensions. Though MR technology has been investigated pervasively, there is no one-size-fits-all definition of MR. In this review, to avoid confusion over the definition of MR, we identified only the articles involving MR technology whose authors explicitly referred to the terms “mixed reality” or “MR”.

Although immersive technologies are not new technologies emerging in educational settings, the rapid development of immersive technology applications in recent years — in terms of interaction capabilities \citep{Petersen2022} and behavioral transition \citep{Miller2016} — have made immersive technologies increasingly attractive to researchers, organizations, and educators. 
With the high level of interactivity, a growing number of mature commercial platforms embedding immersive technologies have become available to domestic consumers, enabling researchers to develop effective pedagogical systems for investigating the interaction behaviors in immersive learning environments at much lower costs. 
Educational scholars from various educational contexts, such as arts \citep{Chang2014}, engineering \citep{Chen2020}, manufacturing and construction \citep{Wu2020}, and science \citep{Cai2021,Chiang2014,Zhang2021}, have made many endeavors to identify behavioral patterns from interactions among learners, teachers, and learning environments during immersive learning processes.
Fruitful behavioral analysis techniques integrating qualitative and quantitative analysis methods have been introduced to the investigation of behavioral processes and uncovering of new research avenues \citep{Hou2012a, Hou18, Lamsa2021}. The behavioral analysis techniques commonly used to construct behavior patterns are summarized in Table \ref{Table1}.
	
\begin{table*}[!htbp]
	\renewcommand{\arraystretch}{1.5} 
	\caption{Behavioral analysis techniques and the associated definition.} 
	\label{Table1}
	\centering
	\scriptsize
	\begin{tabularx}{\textwidth}{p{3cm} p{9cm} X}
		\hline
		\textbf{Behavioral analysis techniques}               &  \textbf{Definition}   &        \textbf{References}\\
		\hline
		Behavior frequency analysis   &  Behavior frequency analysis performs the statistical analysis on the log of the coded behaviors recorded in the interaction system to obtain the behavior’s frequency and distribution information.
		&  \cite{Hou2012b}, \cite{Liu2017} \\
		
		Quantitative content analysis (QCA) &  QCA is a research method defined as systematically, objectively, and quantitatively assigning communication content to categories according to specific coding schemes and rules, and using statistical techniques to analyze the relationships involving these categories.
		&\cite{Poldner2012}, \cite{Riff2014}\\
		
		Lag sequential analysis (LSA)  & LSA a research method that is more appropriate for analyzing the dynamic aspects of interaction behaviors according to time and present sequential chronology information of the users' activities.
		& \cite{Bakeman2011}, \cite{Draper2019}, \cite{Hou09}\\
		
		Social network analysis (SNA)     & SNA is an effective quantitative analytical method for analyzing social structures between individuals in social life, which takes as the origin point the premise that social life is constructed primarily by nodes (e.g., individuals, groups, or committees), the relations between those nodes, and the patterns generated by those relations.
		&  \cite{Scott2011}, \cite{Wu2021}\\
		
		Cluster analysis&Cluster analysis classifies data to form meaningful data groups based on similarity (or homogeneity) in describing the data objects and the relationships among data. 
		& \cite{Tan2019}
		\\
		\hline
	\end{tabularx}
\end{table*}

As highlighted by \cite{Cheng2013}, additional study is necessary to acquire an in-depth understanding of students’ learning behavior sequences by mixed method analysis, such as behavioral analysis techniques, when involved in science education with immersive technologies.
A conceptual framework is proposed to support the development and implementation of learning behavior patterns construction in immersive contexts to inform scholars and researchers investigating the potential pedagogical affordability in relevant educational research.
Additionally, considering review studies that comprehensively explain user behavior patterns analysis in immersive learning environments are still scarce, there is an urgent need to conduct a systematic review to scrutinize users’ behavioral patterns with immersive technology as a whole, looking to explore a deeper understanding of how learners learn and how teachers teach in immersive learning contexts.
Underpinning the proposed conceptual framework, this review contributes to the literature by consolidating factors of behavioral analysis in immersive contexts.

\section{Conceptual framework}

Numerous well-established educational frameworks have been incorporated into immersive learning applications. This section highlights several representative frameworks that form the fundamental basis for the proposed framework.

\cite{Fowler2015} proposed the \emph{design for learning} framework to extend and enhance the pedagogy of immersive learning, which focused more on the pedagogical requirements instead of solely emphasizing technical affordances implicit in 3D virtual learning environments (VLEs). The \emph{design for learning} framework is a practitioner-orientated model to provide guidance to practitioners on designing appropriate 3D VLEs so as to meet the particular teaching and learning requirements. In Fowler’s framework, the learning objectives perform an alignment between learning stages and learning activities, resulting in a complete model covering the stronger pedagogical input and design emphasis for immersive learning environments. This alignment is an attempt to achieve the intended learning outcomes, as described by \cite{Biggs2011}, to acknowledge and understand what competence students are intended to develop. Bloom’s taxonomy \citep{Anderson2001, bloom1956taxonomy} is chosen as the theoretical model to describe the appropriate learning objectives in generic learning activities. How to design immersive learning systems with the learning objectives of analyzing learners’ behaviors in 3D VLEs remains a challenge. Therefore, Fowler’s design for learning framework has a promising potential to provide an avenue for designing effective 3D VLEs with the particular learning requirement of behavioral analysis and provides the foundation for conducting the systematic review in the context of this study.

Another mature conceptual framework used to support the immersive learning system design and development is the \emph{four-dimensional framework} (4DF) \citep{DeFreitas2009, DeFreitas2010,DeFreitas2009b}. The framework comprises the following four dimensions: learner specification, pedagogic perspective, representation, and context. Each dimension of the 4DF has a dependency relationship with others. Meanwhile, the four dimensions would also jointly constitute a robust conceptual framework.

\begin{figure*}[!h]
	\includegraphics{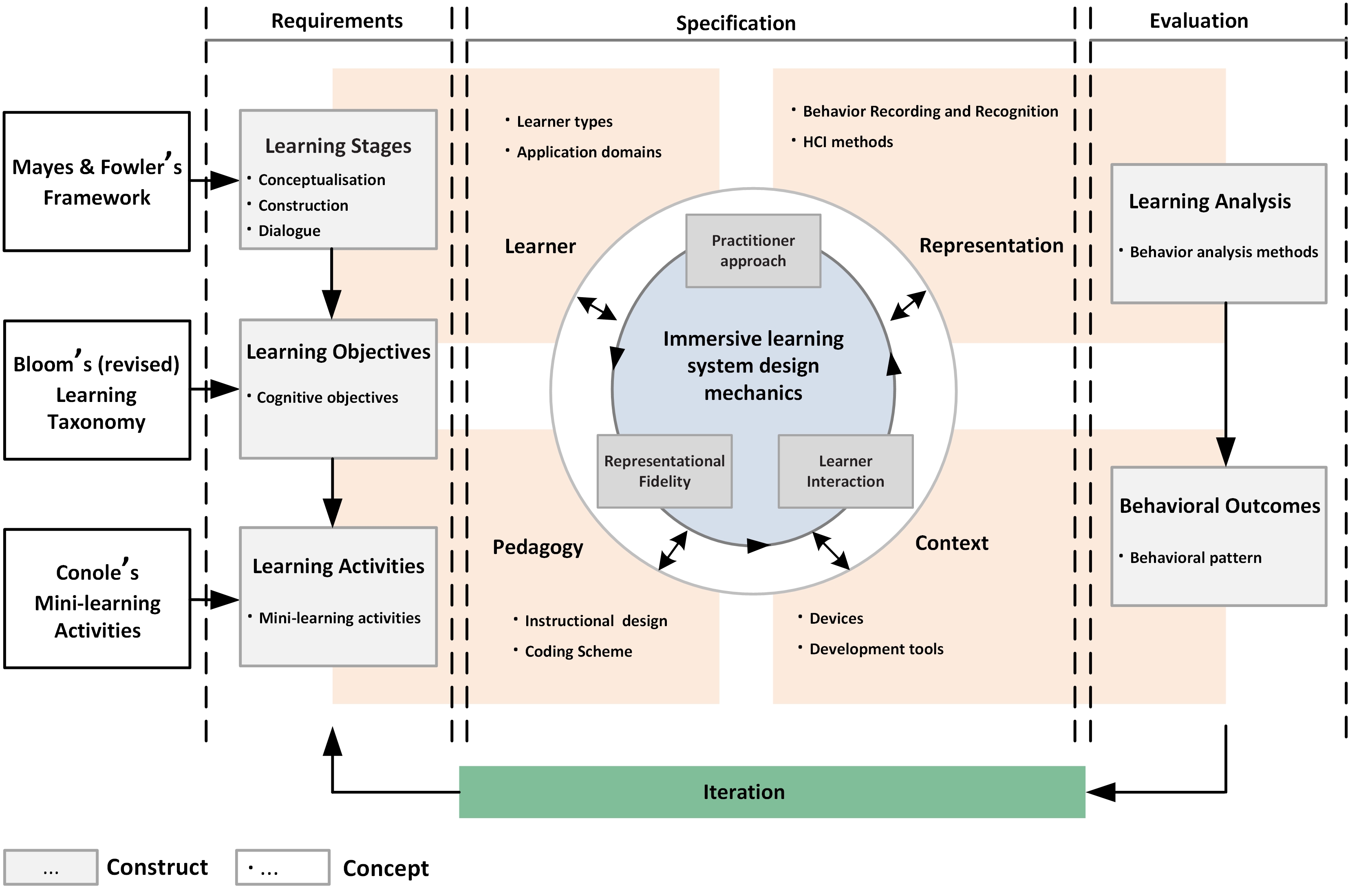}
	\caption{\textbf{Behavioral analysis in immersive learning framework (BAILF)}}
	\label{Fig1}
\end{figure*}

However, one of the critiques of these models and frameworks discussed above is that, more often than not, only a limited number of generic concepts are considered when designing and evaluating immersive learning activities. The high-level models like Fowler’s \emph{design for learning} framework and 4DF are not easy to use for exploratory or explanatory investigation \citep{Mayer2014}, for example, as straightforward theoretical models to review the use and design of immersive learning. 
Poor or sensationalist implementation of learning technologies not restricted to immersive technologies, would seriously hinder the affordability of a promising technology that is applied in the educational domain \citep{Lai2022}. In response to this limitation, a conceptual framework is still needed to provide comprehensive, concrete, and practical perspectives for conducting a systematic review of immersive learning with the intended learning outcomes of behavior analysis in this study. 
Given the benefits as well as the critiques of the discussed frameworks, a behavioral analysis in immersive learning framework (BAILF) is proposed as the extended model of Fowler’s design for learning framework and 4DF that integrates key concepts outlined in previous models to illustrate the requirements, specification, evaluation, and iteration of immersive learning with behavior analysis as the primary intended learning outcomes, as Fig \ref{Fig1} depicts.

The BAILF proposed in this study not only has to be contextualized with relevant literature identified in the following sections, but also needs to consider the higher and lower level construct into the framework. This framework is built upon several pedagogical models, such as activity theory (“four-dimensional framework”), constructivism (“design for learning framework”), and game-based learning framework (“input-process-outcome”, IPO framework) \citep{Garris2002}, and can be summarized into four key stages: requirement, specification, evaluation, and iteration.

\subsection{Requirement.}
The requirement construct is derived from the \emph{design for learning} framework, which envisages the learning design process by satisfying a series of instructional requirements through acknowledging which learning stage the learner is at, setting appropriate learning objectives the learner means to achieve, and conducting a given set of learning activities \citep{Fowler2015}. The requirement construct is treated as the input of the BAILF. However, the \emph{design for learning} framework provides few indications of how to make design decisions on applying the model at the specification stage.

\subsection{Specification.}
In an attempt to overcome this limitation, the 4DF is integrated into the BAILF as the specification stage, promising to encapsulate multiple conceptual theories and frameworks parallelly to increase notable strengths for immersive learning system design and implementation \citep{DeFreitas2013}. The 4DF in the specification stage is rearranged in the following order: \emph{learner}, \emph{pedagogy}, \emph{context}, and \emph{representation}. Furthermore, during the design process of the immersive learning environment, it is crucial to carefully consider two distinctive and unique attributes of 3D Virtual Learning Environments (VLEs): "representational fidelity" and "learner interaction," as emphasized by \cite{Dalgarno2010}, and the specific relationships between these two distinct characteristics should be investigated through appropriate practitioner approaches \citep{Fowler2015}. Consequently, the immersive learning system design mechanics, including three factors: representational fidelity, learner interaction, and practitioner approach, are constructed as the core of the specification stage.
\begin{itemize}
	\item[-] The 1st Dimension in the framework involves the learner specification. The learner-centered construct highlights the significance of learner interaction, one of the two unique characteristics of 3-D VLEs argued in Dalgarno and Lee’s model \citep{Dalgarno2010}. 
	
	\item[-] The 2nd Dimension in the framework analyses the pedagogic perspective when conducting learning activities, and includes a deliberation of instructional models to scaffold learners throughout the learning processes. The selection of learning theories may particularly affect analyzing the intended learning outcomes. Consequently, the systematic review of pedagogic perspectives, such as instructional models and learning theories from related literature, can make sense to find effective ways for knowledge construction and transformation in 3D VLEs. 
	
	\item[-] The 3rd Dimension in the framework outlines the representation of the immersive learning system, including the interactive representation of the learning experience, immersion degree of 3D VLEs, and the representational fidelity, which is the second unique characteristic of 3-D VLEs argued in the Dalgarno and Lee’s model \citep{Dalgarno2010}. 
	
	\item[-] The 4th Dimension in the framework pays attention to the context where the immersive learning happens. Hardware and software platforms are critical factors for supporting the construction of the learning context.
\end{itemize}

\subsection{Evaluation.}
Inspired by the IPO framework, the model's output construct is vital for evaluating the achievement of learning objectives and intended learning outcomes \citep{Ak2012, Garris2002,Hsiao2016}. To this end, the evaluation construct is integrated into the BAILF, which comprises three factors: learning analysis methods and behavioral outcomes.

\subsection{Iteration.}
To make the learning system scalable and robust, the iteration stage as the refining process in framework construction is important for preventing sensationalist or haphazard integration of immersive technology into learning practices \citep{DeFreitas2006}.
In the iteration stage, problems and limitations in the system design and evaluation process are discovered, and additions, subtractions, as well as substitutions of learning system implementation should be carried out to fix the system defects.

\section{Research method: systematic literature review}

Aiming to achieve research objectives of the systematic literature review, this systematic review followed the preset protocol to provide essential both quantitative and qualitative evidence \citep{Arksey2005,Khan2003,Wendler2012}.

\subsection{Information sources}

The papers were recovered from the online research databases related to education and technology, including Scopus, Web of Science, IEEE Xplore Digital Library, and ERIC (Education Resources Information Center). This systematic literature review also was mindful of other online databases, such as Science Direct, the ACM digital library, JSTOR, Wiley, EBSCO, and Taylor \& Francis. Nonetheless, the majority of relevant articles retrieved from these datasets were anticipated to be already included in the datasets selected above, which has been confirmed using exemplary cross-checks. To avoid omitting some relevant papers that were not included in these databases, the backward and forward snowballing method, which is a practical literature searching technique for identifying additional relevant papers based on the references list and the citations of the target papers \citep{Wohlin2014}, was applied in the literature identification procedure. StArt (state-of-the-art through systematic review) tool \citep{Fabbri2016} was adopted as information extraction software to assist in data organization and monitoring to decrease the chances of errors when processing duplicate papers.

\subsection{Search criteria}

The search terms or keywords include any query strings that describe immersive technologies and additional query strings that describe the construction of behavioral patterns by utilizing behavioral analysis techniques in an instructional context. The research terms used for literature identification are summarized in Table \ref{Table3}. According to the characteristics of each online research database, the search strings were manually combined and marginally modified to match the search capabilities provided by each database.

\begin{table*}[!htbp]
	\renewcommand{\arraystretch}{1.5} 
	\caption{Key search terms used for literature identification that was conducted in November 2022.}
	\label{Table3}
	\centering
	\scriptsize
	\begin{tabularx}{\textwidth}{p{4cm} X p{6cm} X p{3cm} }
		\hline
		\textbf{Immersive technology-related concept}  & AND & \textbf{Behavioral analysis-related concept} & AND & \textbf{Education-related concept}\\
		\hline
		Immersive technologies* OR
		
		Virtual reality* OR VR OR
		
		Augmented reality* OR AR OR
		
		Mixed reality* OR MR OR
		
		Cross reality* OR
		
		Extended reality* OR XR
		& &
		Behavior* analysis OR
		
		Behavioral pattern* OR 
		
		Quantitative content analysis* OR QCA OR
		
		Lag sequential analysis* OR LSA OR
		
		Social network analysis* OR SNA OR
		
		Cluster analysis		
		&  &
		Education* OR 
		
		Learn* OR 
		
		Train* OR 
		
		Teach* OR
		
		Student*		
		\\
		\hline
	\end{tabularx}
\end{table*}

\subsection{Inclusion and exclusion criteria}

\begin{figure*}[!h]
	\includegraphics{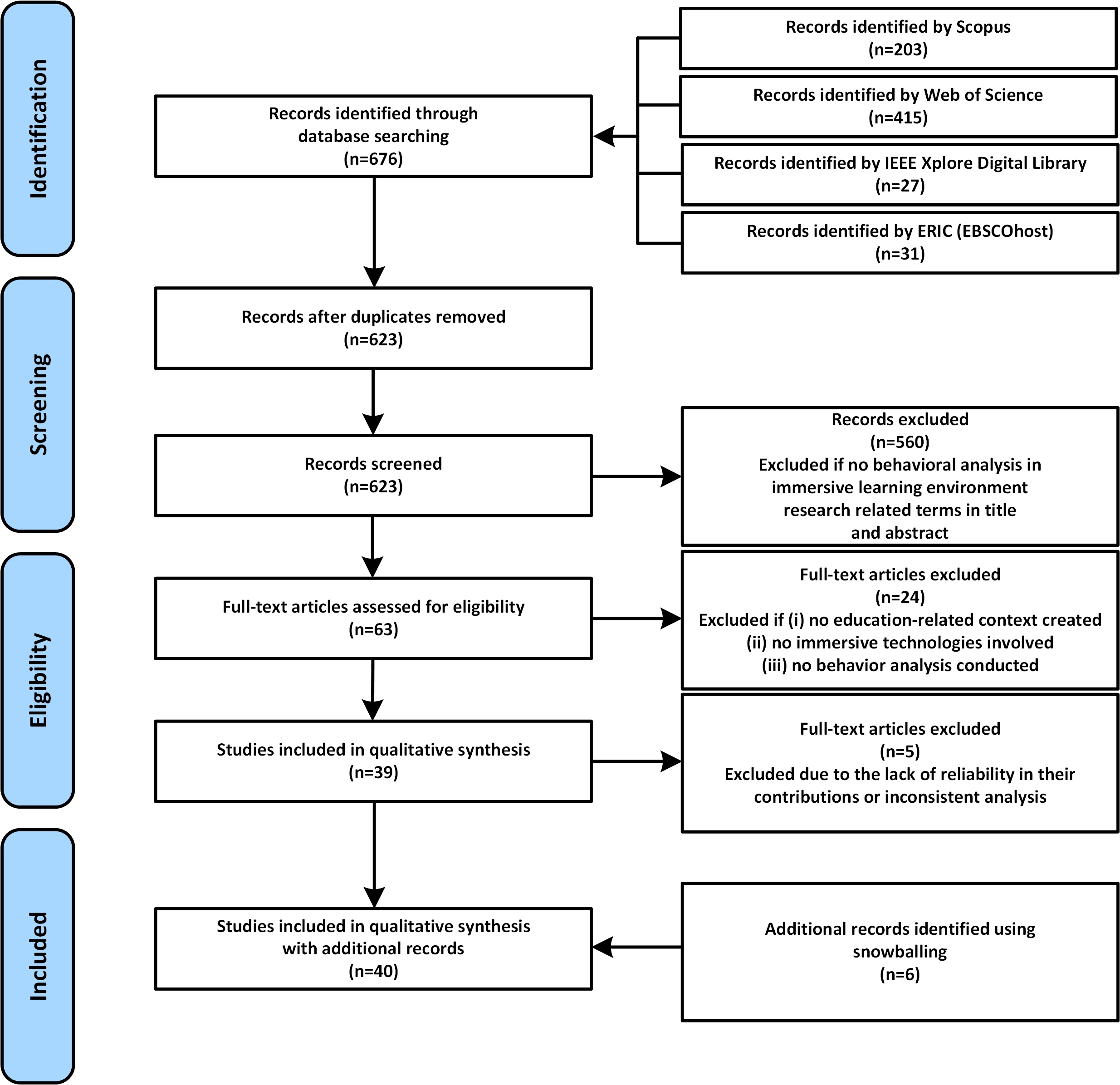}
	\centering
	\caption{
		\label{Fig2}
		\textbf{Literature identification process derived from the PRISMA framework}. Using the previously-defined key concept terms, this review yielded 676 results from databases. With the aid of the StArt software, 53 duplicate papers were found and deleted. During the screening phase, a review of the titles, abstracts, and keywords revealed 560 irrelevant articles, and 63 studies that met the inclusion criteria were included for the following selection stage. During the eligibility phase, the entire text of the remaining articles was scanned in detail to check the theoretical contribution in the area of learning implementation, behavior analysis, virtual communities, and the extension of instruction theories. Thus, 24 studies with limited evidence on analyzing learner behavior sequences or without immersive intervention were excluded. Additionally, another five articles were classified as ineligible due to the lack of reliability in their contributions or inconsistent analysis. Then, the backward and forward snowballing method was carried out on Google Scholar to find more relevant literature, including another six papers in this review. In the end, 40 papers were ultimately classified as eligible and included in the final review.}
\end{figure*}

The paper selection procedure was guided by the principles of the Preferred Reporting Items for Systematic Literature Reviews and Meta-Analyses (PRISMA) \citep{Moher2009} statement, as shown in Fig \ref{Fig2}. To determine whether a study meets the eligibility criteria, inclusion and exclusion criteria aimed at addressing research questions were proposed during the screening stage.

\emph{Inclusion criteria.}
\begin{enumerate}[label=(\arabic*)]
	\item {Papers were peer-reviewed primary source articles.}
	\item Papers whose full text was accessible.
	\item Research areas: education-related, immersive technology-related papers.
	\item Papers dealt with behavioral patterns analysis and construction.
\end{enumerate}

\emph{Exclusion criteria.}
\begin{enumerate}[label=(\arabic*)]
	\item {Papers were commentaries, literature reviews or book chapters.}
	\item {Papers whose full text was not accessible.}
	\item {The focus of the study was not on the education-related, immersive technology-related context.}
	\item {The analysis of the study did not involve the content about learners’ behavior analysis.}
\end{enumerate}

\subsection{Literature quality assessment}

Aiming to evaluate the quality of eligible papers, a series of quality criteria derived from the literature quality assessment method presented by \cite{Connolly2012} was applied and revised in this study.

Specifically, each of the included papers was analyzed and assigned a score ranging between 1 to 3 across the five dimensions, where 1 indicates low quality, 2 indicates medium quality, and 3 indicates high quality in each dimension. Each paper was graded on five dimensions, with the sum of the scores for each dimension determining the final score. The paper quality assessment procedure was conducted independently by two raters, and the final score for each paper was determined by computing the mean value of the two raters’ scores, ranging from 5 to 15.  As suggested by \cite{JOHNSON201689}, the articles could be classified into three categories according to the scores: articles with less than 8 points were annotated as “weaker evidence”, articles with 8-12 points were annotated as “moderate evidence”, articles with more than 12 points were annotated as “stronger evidence”.

\begin{table*}[!h]
	\renewcommand{\arraystretch}{1.2}
	\caption{Classification scheme.}
	\label{Table2}
	\centering
	\scriptsize
	\begin{tabularx}{\textwidth}{p{3cm} p{3cm} X }
		\hline
		\textbf{Concept matrix facets}               &  \textbf{Categories}   &        \textbf{Description}\\
		\hline
		
		\multirow[t]{3}{3cm}{1.1 Learning stages}&Conceptualization	&Learners come into contact with concepts through presentation and visualization in the immersive learning environment.
		\\									&Construction		&Learners construct new knowledge through interactivity with others or virtual learning content in the immersive learning environment.
		\\									&Dialogue			&Learners test their emerging understanding of new knowledge through discussion with others or a more comprehensive range of interactivity in VLEs.
		\\
		
		\multirow[t]{2}{3cm}{1.2 Cognitive learning outcomes/objectives}
		&Lower-level cognitive category	& Remembering is the cognitive process with low cognitive complexity, including identifying and recalling relevant information from long-term memory.
		
		Understanding is the cognitive process that helps learners construct meaning from instructional activities and has subcategories such as interpreting, exemplifying, classifying, summarizing, inferring, and explaining \citep{Radmehr2019}.	
		\\&Higher-level cognitive category & Applying is to implement the acquired knowledge into practice.
		
		Analyzing is to break the learned knowledge into constituent parts and determine the relationship of the parts with overall structure.
		
		Evaluating is to judge the learned knowledge based on specific criteria.
		
		Creating is to make new learning products by mentally reorganizing fragmented elements into new knowledge patterns or structures.		
		\\
		
		1.3 Learning activities&List of mini-learning activities&Learning activities are the actions learners display to reach the intended learning goals. Individual mini-learning activities (behaviors) can be grouped into learning activities at an extensive range of granularity through specific behavioral patterns.
		\\
		
		2.1 Learner types&Specific learner types&The specific type of learners when participating in the immersive learning activities.
		\\
		
		\multirow[t]{3}{3cm}{2.2 Application domains}
		&STEM	&Science, Technology, Engineering, and Mathematics (STEM) describes various academic disciplines related to these four terms, such as: biology, chemistry, engineering, mathematics, physics, and more.
		\\									
		&Humanities	&Humanities describe academic disciplines that study aspects of human society and culture, including culture, history, language, and more.
		\\									
		&General Knowledge \& Skills	&General knowledge \& skill describes the application domains where learners study basic knowledge and skills to cultivate the essential ability to deal with daily affairs, such as cognitive \& social skills, art \& design, and reading.
		\\
		
		\multirow[t]{2}{3cm}{3.1 Instructional design methods}
		&Instructional Strategies& The instructional strategies entail a set of instructional models to lead learners to understand what information has been provided, how the learning process functions, and how to acquire learning acquisition effectively. 
		\\&Instructional Techniques & The instructional techniques are the rules, procedures, tools, and skills used to implement the instructional strategies into practice. 
		\\
		
		3.2 Coding schemes&Specific Coding schemes&The coding scheme defines the specific behavior sequences that would be analyzed using various behavioral analysis techniques.
		\\
		
		4.1 Hardware devices&Specific Hardware devices&Hardware devices used in the immersive learning activities.
		\\
		
		4.2 Software development tools&Specific software explicitly&Software tools used to develop immersive learning systems.
		\\
		
		5.1 HCI methods in VLEs&Specific HCI methods&The interaction methods between learners and VLEs.
		\\
		
		\multirow[t]{2}{3cm}{5.2 Behavior recording and recognition methods}
		&Specific behavior recording methods& The applied methods used to record learners’ behavior sequences, such as videotaping, classroom observation, and automatic recording methods by software tools.
		\\&Manual and automatic coding methods &The manual coding method refers to the method of behavior recognition that is conducted manually and usually independently by two or more coders.
		
		Automatic coding methods refer to behavior recognition that is conducted using software tools automatically.
		\\
		
		6.1 Behavioral analysis methods&Specific Behavioral analysis methods& Behavioral analysis methods are used to analyze learners’ behavior sequences to construct behavioral patterns, such as behavior frequency analysis, QCA, LSA, SNA, and cluster analysis.
		\\
		6.2 Behavioral patterns outcomes&Constructed behavioral patterns&Behavioral patterns are constructed as the vital outcome of behavioral analysis in immersive learning environments.
		\\
		7.1 learning iteration	&Iteration expectation and difficulties&The learning iteration requirements were found in the implementation of learning activities of behavioral analysis in immersive learning environments.
		\\
		\hline
	\end{tabularx}
\end{table*}

\subsection{Coding Procedure}
With the aim of conducting a systematic review mapping study on the theme of behavioral analysis in the immersive learning environment,  it is essential to elaborate every construct in the BAILF, which produces problems that need to be addressed corresponding to the framework constructs.  In order to obtain a deeper understanding of the generalization of BAILF, a concept matrix adapted from \cite{Salipante1982} and \cite{Webster2002} was developed to make the transition from the author- to the concept-centric literature review, providing classification structure in helping clarify critical concepts of the review. Seven main concept matrix facets correlated with the relevant framework constructs were formulated, as summarized in Table \ref{Table2}. Based on the classification scheme, seven primary research questions were proposed in this review:

\begin{itemize}
	
	\item[-] RQ1. What are the learning requirements in the design of behavioral patterns construction in immersive learning environments?
	
	\item[-] RQ2. What are the learner specifics based on 4DF for behavioral analysis in immersive learning environments?
	
	\item[-] RQ3. What are the pedagogic considerations based on 4DF for behavioral analysis in immersive learning environments?
		
	\item[-] RQ4. How to construct immersive context based on 4DF for behavioral analysis?
			
	\item[-] RQ5. What are the representation dimension based on 4DF for behavioral analysis in immersive learning environments?
	
	\item[-] RQ6. What behavioral patterns were constructed in  immersive learning environments?
	
	\item[-] RQ7. What are the challenges in analyzing learners' behavior in immersive learning environments?
	
\end{itemize}

\section{Results and analysis}

In this section, the evidence extracted from the retained literature based on the above work is analyzed and interpreted from the four stages of BAILF, providing detailed, structured, and concrete information on the themes of this review. All included papers were coded according to the major immersive technology involved, and the behavioral analysis-related literature evidence was also summarized, as shown in Table \ref{TableA1}. Accordingly, the answers to formulated research questions were revealed and presented.

\begin{figure*}[!h]
	\includegraphics{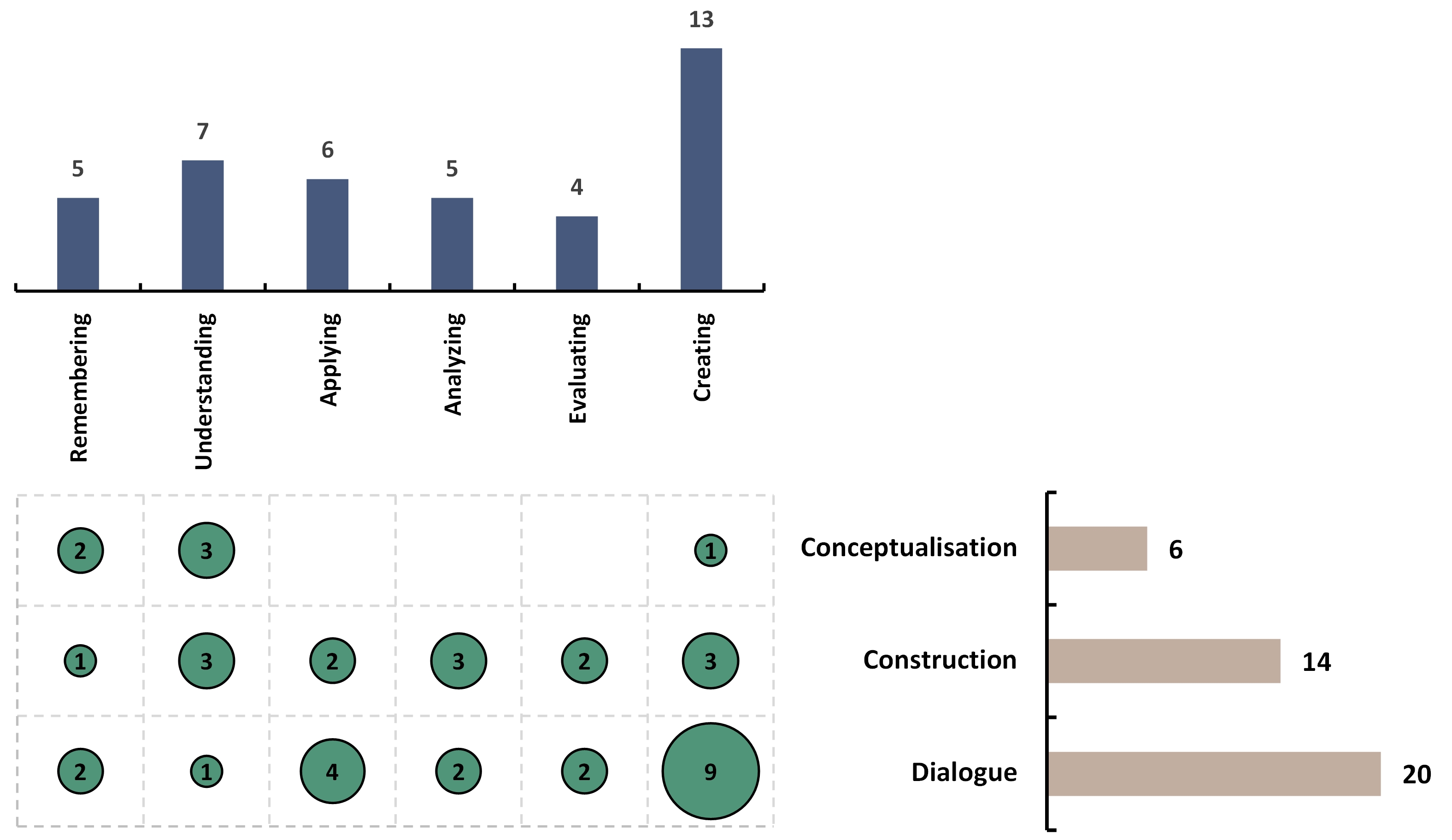}
	\centering
	\caption{
		\label{Fig3}
		\textbf{Learning stages and cognitive learning objectives/outcomes}. In the bottom right bar chart, a majority of researchers have implemented immersive learning for the higher echelons of learning stages, i.e., dialogue (n=20), followed by construction (n=14) as the mediate echelons of learning stages, and conceptualization (n=6) as the lower echelons of learning stages. In the upper-left bar chart, the most common cognitive learning outcome based on Bloom’s revised taxonomy was the ability to create knowledge (n=13). Remembering (n=5), understanding (n=7), applying (n=6), and analyzing (n=5) were the following most common cognitive learning outcomes that participants achieved in the immersive learning environments. Finally, in 4 studies, participants acquired the ability to evaluate the immersive learning activities.}
\end{figure*}

\subsection{RQ1. What are the learning requirements in the design of behavioral patterns construction in immersive learning environments?}

Based on the conceptual frameworks from design for learning \citep{Fowler2015} and the proposed BAILF, three constructs are selected to uncover the learning requirements in planning behavioral patterns construction: learning stages, cognitive learning objectives/outcomes, and learning activities (see 1.1-1.3 in Table \ref{Table2}).

To look deeper into the relationship between the learning stages and learning objectives in the learning requirements, this review adapted the two broad categories of cognitive learning outcomes from \cite{Ibanez2018}, who classified the measured cognitive outcomes into two broad categories based on the revised Bloom’s Taxonomy: the lower-level cognitive category covered works dealing with simpler cognitive processes including remembering and understanding; the higher-level cognitive category covered works dealing with more complicated cognitive processes, including applying, analyzing, evaluating, and creating. A bubble chart to represent the concentration of the studies was prepared in the bottom left part of Fig \ref{Fig3}. The bubble chart shows that the majority of studies with the higher echelons of the learning stage “dialogue” are possible, which also allow learners to “create” (n=9), “apply” (n=4), “analyze” (n=2), and “evaluate” (n=2) the learned knowledge, indicating that the dialogue learning stage depends on the learners' ability to have a deeper understanding of the concepts learned, and to carry out structured thought from debates and discussions to reflect the expert knowledge, and to “identify” the subject matter \citep{Fowler2015,Mayes1999}. Conversely, learners in the lower echelons of conceptualization learning stages may only acquire the ability to remember (n=2) or understand (n=3) the learned concepts at the lower cognitive level. Interestingly, the bubble chart shows that in those studies with construction learning stages where learners have the essential ability to control the flow of learning information through learning interactives, the immersive learning systems can provide learners with affordances for obtaining not only higher cognitive learning outcomes (n=10), but also lower cognitive learning outcomes (n=4).

The mini-activities are matched to the learning stages. The short lists of typical Mini-activities (see 3.1 in Table \ref{Table2}) retrieved from identified articles were tabulated in Table \ref{TableA1}. In those 6 articles that only implemented immersive learning at the primary conceptualization stage, researchers designed mini-activities to provide a primary exposition of the concept to be formed, allowing learners to be immersed in the concept representation using various immersive technologies. Though the primary exposition can only serve to provide the superficial initial contact with the conceptual knowledge to be learned, the essential function in the primary conceptualization stage is to orient the learners and give learners clear learning maps onto subject matter through appropriate learning activities \citep{Mayes1999}, such as constructing environments, observing models, receiving information, and discovering facts. Based on the primary conceptualization stage for representing knowledge, researchers can design task-based mini-activities in the secondary construction stage (n=14), where learners can engage at a higher conceptual level through experiential and contextual learning \citep{Lai2022}. In the tertiary dialogue stage (n=20), learners’ developing understandings need to be tested through elaborated mini-activities, such as reflective thinking with themselves, synchronous or asynchronous discussion with peers, and real-time collaborative learning.  The findings suggest that researchers and designers should specify representative learning activities to evaluate which stage the learner stays in.

\begin{figure*}[!h]
	\includegraphics{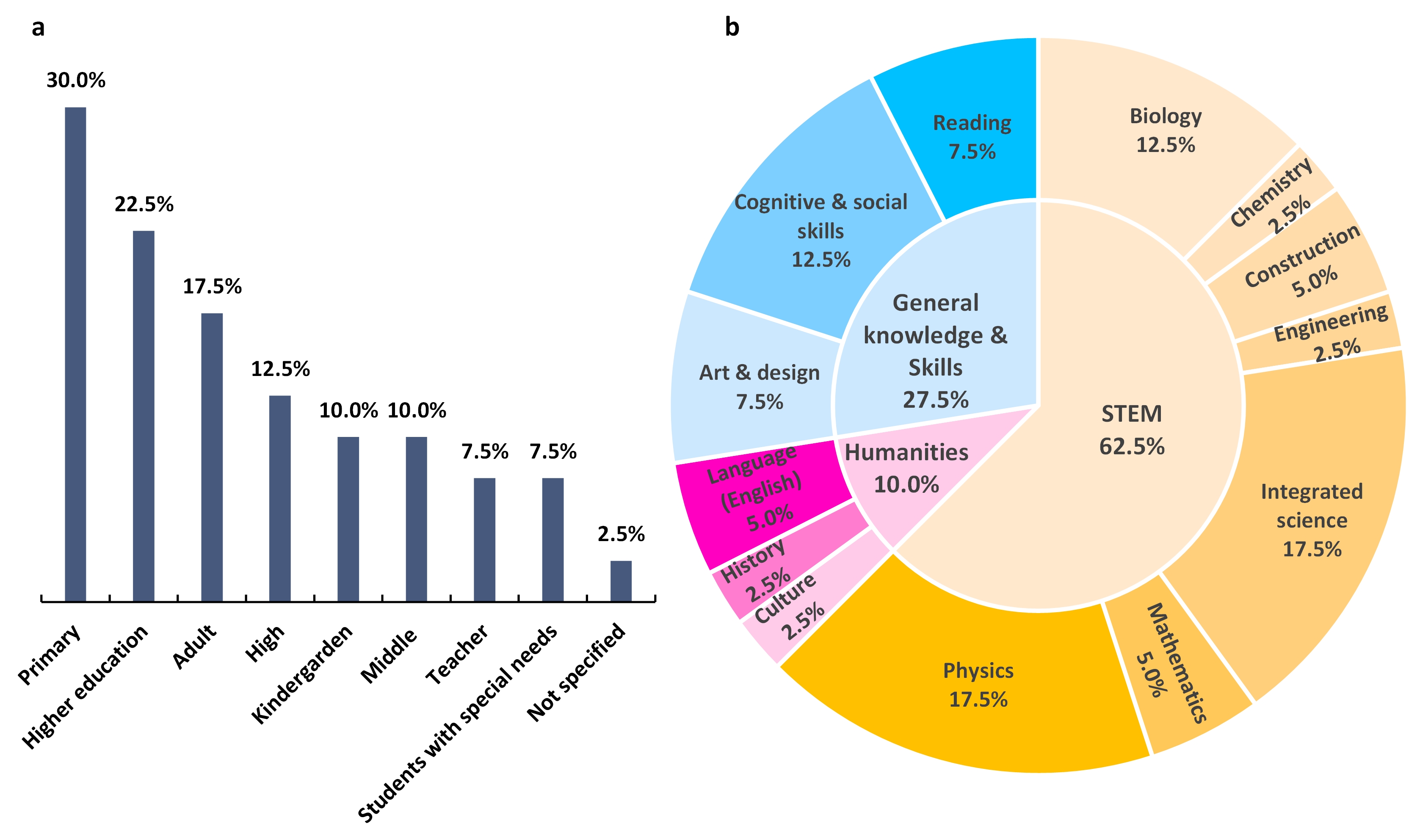}
	\centering
	\caption{
		\label{Fig4}
		\textbf{Learner specifics}. \textbf{a},  learner types. The study participants were mostly primary school students, with 12 papers accounting for 30\% of the total, and higher education students, with 9 papers accounting for 22.5\% of all papers. Other studies tended to recruit learners of adults (17.5\%, n=7), high school students (12.5\%, n=5), kindergarten children (10\%, n=4), middle school students, teachers (7.5\%, n=3), and students with special needs (7.5\%, n=3). One paper, accounting for 2.5\% of the total, did not specify the learner type in the article content. \textbf{b}, application domain. More than half of the articles chose STEM (62.5\%) as the application domain of their learning systems. The second popular application domain was general knowledge \& skills (27.5\%), where learners can learn basic social or art abilities to deal with daily affairs. The rest of the articles chose to learn about knowledge in the humanities (10\%). Specifically, in the category of STEM, physics (17.5\%), integrated science (17.5\%), and biology (12.5\%) were popular topics in the immersive learning system. “Integrated science” refers to the specific subject that involves more than one scientific discipline in the learning activities. In the humanities category, history (2.5\%), culture (2.5\%), and language learning (5\%) were the common topics. In the general knowledge \& skills category, the literature highlighted three topics of educative applications: art \& design (7.5\%), cognitive \& social skills (12.5\%), and reading (7.5\%).}
\end{figure*}

\subsection{RQ2. What are the learner specifics based on 4DF for behavioral analysis in immersive learning environments?}

The learner specifics dimension specified two concrete research factors about the population information of learners: learner types and application domains (see 2.1-2.2 in Table \ref{Table2}).

Fig \ref{Fig4}(a) presents the percentage of specific learner types across all articles. Since there may be more than one type of learner involved in the learning activities, such as children with their parents \citep{Cheng2014,Cheng2016,Hsu2020}, students with teachers \citep{Cai2021,Yilmaz2016}, or students with other adults \citep{Wu2019}, the sum of all percentages were greater than 100\%. The largest proportion of learners, comprising 52.5\%, belonged to the K12 category, which includes primary, middle, and high school students. This finding suggests that research tend to use immersive technology to enable young learners to experience media-rich virtual environments. Furthermore, given that children nowadays tend to spend a considerable amount of time playing electronic games \citep{Lee2012}, immersive learning experiences in the form of digital games continue to remain appealing to young learners.  In highly interactive immersive learning environments, it is possible to generate, observe and track abundant learning behaviors. In-depth analysis of these behaviors can provide vital insights into the cognitive and affective processes of learners, the levels of engagement, and the learning outcomes.

The application domains identified in the articles were firstly categorized into three broad categories: STEM, humanities, and general knowledge \& skills, as shown in Fig \ref{Fig4}(b). Subsequently, 13 sub-categories were created to further refine the learner application domains.  This result suggests that researchers have conducted experiments to analyze leanrers' behavior in immersive learning environments across broad application domains. The popularity of STEM as a major domain in immersive technology-based learning systems is consistent with the findings of \cite{Law2021}.

 It should also be noted that the only three papers that have addressed the targeted autistic learners with special needs have all decided to teach learners cognitive \& social skills to help them deal with social relationships with other people.

\subsection{RQ3. What are the pedagogic considerations based on 4DF for behavioral analysis in immersive learning environments?}

\begin{table*}[!h]
	\renewcommand{\arraystretch}{1.5} 
	\caption{Summary of instructional design methods.}
	\label{Table4}
	\centering
	\scriptsize
	\begin{tabularx}{\textwidth}{p{2.8cm} X p{9cm} p{2cm} }
		\hline
		\textbf{Instructional design methods}               &  \textbf{Categories}   &        \textbf{Description}& \textbf{Paper code}\\
		\hline
		
		\multirow[t]{5}{2.8cm}{Instructional Strategies}
		&Presentation	&Presentation is the instructional strategy that suggests learners get new knowledge through the presentation of learning tasks or material to strengthen cognitive organization \citep[ pp. 65]{Akdeniz2016}. &	A2, A4, A18, A19, A20, V2, V8, M1
		\\								
		&Discovery &	Discovery is the instructional strategy that suggests learners get new knowledge through discovering rather than being told about the information \citep[ pp. 65]{Akdeniz2016}.&	A8, A9, V3, V5, V6, V7, M5		
		\\									
		&Inquiry&	Inquiry is the instructional strategy that emphasizes that learners actively participate in the learning process, where the learners’ inquiries, thoughts, and observations are placed as the focal spot of the learning process \citep[ pp. 67]{Akdeniz2016}.&		A1, A10, A13, A15, V1, V4, V13, V14, M4		
		\\
		&Collaborative&	Collaborative is the instructional strategy that suggests learners get new knowledge through working in a social setting to solve problems \citep[ pp. 68]{Akdeniz2016}.&		A3, A5, A7, A11, A14, A21, V10, V11, V12, M2, M3
		\\
		&Collaborative Inquiry	&Collaborative Inquiry is the instructional strategy that suggests learners conduct scientific inquiry learning through face-to-face collaboration \citep{Chiang2014,Wang2022}.& A6, A12, A16, A17, V9
		\\

		\multirow[t]{6}{2.8cm}{Instructional Techniques}
		&		Observation&	Observation technique is the instructional technique that suggests learners monitor and examine the indicators or conditions of objects, facts or materials within a well-designed plan through eyes or available visual equipment \citep[pp. 204-205]{Gunduz2016}.& A1, A2, A3, A4, A8, A9, A13, A13, A18, A20, A21, V9
		\\
		&
		Field trip&	Field trip is the instructional technique that suggests learners gain additional knowledge through direct experiences in conducting the active research-oriented field project \citep[pp. 196-198]{Gunduz2016}.& A6, A12, A15, A19, V5
		\\
		&
		Educational Game&	Educational game is the instructional technique that suggests learners gain knowledge through playing educational games to increase learning motivation and promote creative work \citep[pp. 201-203]{Gunduz2016}.&
		A5, A7, V6, V7, V8, M1
		\\
		&
		Role-play	&Role-play is the instructional technique that suggests learners play specific roles in the explicitly established situation and gain knowledge through experiencing their “character” \citep[pp. 172-174]{Gunduz2016}.& A7, V3, V6, V7, V12
		\\
		&
		Simulation	&Simulation is the instructional technique that suggests learners gain knowledge in a controlled detailed situation that intends to reflect real-life conditions \citep[pp. 187-189]{Gunduz2016}.& A10, A11, A14, A16, A17, V2, M3, M4, M5
		\\
		& Project &Project is the instructional technique that suggests learners are involved in whole-hearted purposeful learning activities to accomplish a specific goal \citep[ pp. 198-201]{Gunduz2016}. & V1, V4, V10, V11, V13, V14, M2		
		\\
		\hline
	\end{tabularx}
\end{table*}

The pedagogic considerations dimension specifies two concrete research factors about the studies’ pedagogical and theoretical information: instructional design methods and behavioral coding schemes (see 3.1-3.2 in Table \ref{Table2}).

Table \ref{Table4} tabulates the instructional design methods, which comprise the instructional strategies and instructional techniques according to \cite{Akdeniz2016}. Additionally, the retained literature was analyzed and divided into five sub-categories of instructional strategy: \emph{presentation}, \emph{discovery}, \emph{inquiry}, \emph{collaborative} and \emph{collaborative inquiry}. Accordingly, six sub-categories of instructional technique were used to classify the reviewed literature: \emph{observation}, \emph{field trip}, \emph{game}, \emph{role-play}, \emph{simulation}, and \emph{project}.

Concerning the instructional strategies, eight different immersive applications reviewed followed the presentation (n=8) instructional strategy, presenting the supplementary virtual learning materials to learners using the immersive technologies. Seven studies implemented the discovery (n=7) strategy, allowing learners to construct knowledge in self-directed and constructivist conditions in immersive learning environments. Nine studies followed the inquiry (n=9) strategy, through which learners play more active roles in a series of immersive learning activities, including raising questions, drawing up learning plans, observing phenomena, and solving problems. A solid foundation of pertinent literature evidence (n=11) was guided by the collaborative strategy, indicating that researchers are more willing to plan and implement learning activities through collaborative group work in immersive learning environments. Furthermore, a relatively complex learning strategy named collaborative inquiry (n=5) was adopted to guide five studies, which indicates that learners conduct inquiry learning activities through group work in immersive learning environments.

Regarding the instructional techniques, observation (n=13) was the most often deployed technique in immersive educational studies. Referring to the paper codes of the reference papers that deployed the observation technique, it can deduce that the AR-based studies mostly used observation in the instruction design, enabling learners to observe more superimposed virtual information onto the real objects. The field trip (n=5) technique that provides more active characters for learners outsides the traditional classroom settings has been employed by five studies, including four studies that conducted AR-based learning activities outside the classroom and one study that developed the virtual field trip in VR space supported by 360° panoramic images \citep{Cheng2019}. As a popular instructional technique to bring learning interest and promote learning motivation for learners, educational games (n=6) were used in six studies reviewed. As an instructional technique to support learners immersing in the characters of instructional activities, role-play (n=5) was used in five studies. Nine studies adopted simulation (n=9) as the instructional technique, enabling learners to realize essential physical phenomena in the immersive world that are difficult to come into being in the real world. Finally, the relatively complex instructional technique name project (n=7) was deployed in seven studies.

The coding scheme is vital in learning behavioral analysis because it defines and classifies a specific phase of the behavior sequence. Using the coding scheme, researchers can recognize and code each behavioral message by matching the dominant content of the message with the item defined in the coding scheme best applied to the content of the message through various behavioral manual or automatic recognition methods \citep{Hou2010}. In this review, three broad categories of coding schemes were detected from the identified literature: coding scheme developed by the authors (n=13), coding scheme based on previous literature (n=25), and not specified (n=2), as shown in Table \ref{TableA1}. In those 13 articles where the coding schemes were designed and developed by the researchers themselves, the behavioral items defined in the schemes reflected the structure of the immersive learning activities \citep{Chang2020,Ibanez2016}, corresponding to the mini-activities summarized in Table \ref{TableA1}. The pervasive investigation of behavioral patterns in educational areas has cultivated abundantly available coding schemes, which provided theoretical bases for subsequent studies to improve and adjust new applicable coding schemes, leading to the extensive adoption of such adaptation of existing coding schemes in most studies. Finally, in two studies, the researcher obtained behavioral sequences analysis outcomes using various data processing methods, such as cluster analysis \citep{Cheng2015} and data triangulation technique \citep{Lorenzo2012},  without providing explicit coding schemes.

\subsection{RQ4. How to construct immersive context based on 4DF for behavioral analysis?}

\begin{figure*}[!h]
	\includegraphics{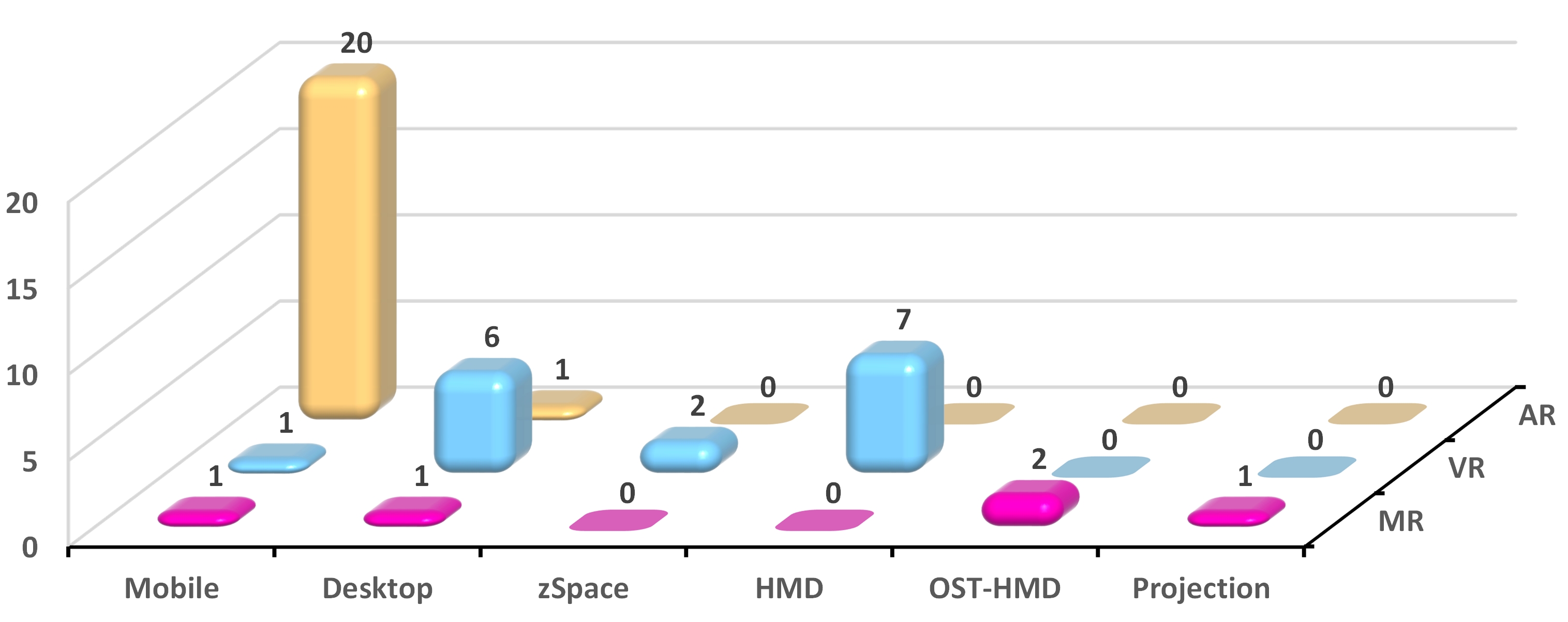}
	\centering
	\caption{
		\label{Fig5}
		\textbf{Hardware devices}. As for AR condition, except for one paper that used desktop computing devices (n=1) to construct the learning system, all other papers used mobile devices (n=20) as predominant apparatus. As for VR conditions, the non-immersive VR was equipped with mobile devices (n=1) and desktop computing devices (n=6). Full-immersive VR that adopted HMD devices (n=7) as infrastructure was widespread in the immersive learning system construction. Two studies used zSpace as hardware devices, which is expected to provide learners with a semi-immersive experience in this review. As for MR conditions, four kinds of hardware devices were used to set up MR systems: mobile devices (n=1), desktop computing devices (n=1), OST-HMD (n=2), and projection-based apparatus (n=1).}
\end{figure*}

Context dimension focuses on the construction the specific environment where the learning process takes place. In this review, the hardware and software used to set up the immersive learning context were analyzed from the literature (see 4.1-4.2 in Table \ref{Table2}).

In relation to the hardware devices that were used to set up the learning systems, categories have been summarized according to the immersive technologies used in the articles, as shown in Fig \ref{Fig5}. 
This result indicates that the widespread adoption of mobile devices as the primary apparatus in most AR learning systems is due to their low cost, high flexibility, and satisfactory performance \citep{Mystakidis2022}. Nonetheless, the uniqueness of the augmented reality (AR) devices considered in the included papers could potentially impact the generalizability of valuable insights into the behavioral effects of general AR technologies.
In VR conditions, the study's findings indicate that the investigation of behavioral patterns in VR environments encompassed the entire spectrum of VR devices, including non-immersive, semi-immersive, and fully immersive systems.
Though the number of studies that used MR technologies to analyze specific behavioral patterns was small, the hardware devices used to set up MR systems were various in this review. Among these hardware devices, the optical see‐through head‐mounted displays (OST‐HMDs) are projected to be more widely used in MR learning research for the advantages of a high degree of interactivity and scalability \citep{Gao2019}.

It is highly significant to observe the numerous software tools available for constructing immersive learning systems in the context of behavioral analysis. The specific software tools utilized in each article were outlined in the supplementary material. Native applications (n=7) which were based on existing commercial applications or program packages developed by previous studies, and self-developed software tools (n=7) were mostly utilized to create immersive learning environments. On the one side, the reason for the popularity of native applications may derive from the convenience and low cost of native applications for educators and researchers who lack advanced programming backgrounds \citep{Cheng2019,Zhang2016}. On the other side, for educators and researchers with proficient programming ability, self-developed software tools ensure that they have more freedom to realize the design features of the learning system and make the system easier to iterate \citep{Lin2013,Lin2022}. Several papers developed their immersive learning systems frequently using Unity 3D (n=5) as the development platform for advanced graphical and visual features as well as flexible cross-platform capability to integrate with extensive software development kit (SDK), such as Vuforia \citep{Cai2021} and ARCore \citep{Sarkar2020}. Eight studies did not specify certain software tools for creating immersive learning environments. Finally, the rest of identified papers tended to use unique software tools for immersive system development resulting in a total of 17 categories in software tool classification, which indicates that there are plenty of software solutions that educators and researchers can leverage to set up the practical immersive systems no matter what degree of programming skills they have mastered.

\subsection{RQ5. What are the representation dimension based on 4DF for behavioral analysis in immersive learning environments?}

\begin{table*}[!htbp]
	\renewcommand{\arraystretch}{1.5} 
	\caption{Summary of behavior recording and recognition methods.}
	\label{Table5}
	\centering
	\scriptsize
	\begin{tabularx}{\textwidth}{p{5.5cm} p{5cm}X  }
		\hline
		\textbf{Behavior recording methods}               &  \textbf{Behavior recognition methods}   &        \textbf{Paper code}\\
		\hline
		
		\multirow[t]{2}{5.5cm}{Videotaping}
		&	Manual coding&	A1, A2, A5, A7, A13, A18, A19, V1, V2, V4, V5, V9, V10, V11, V13, M1, M5
		\\
		&Automatic coding&	A3, A4, A14		
		\\
		
		\multirow[t]{2}{5.5cm}{Videotaping combined with other manual recording methods}
		&Manual coding	&A11, A16, A17\\
		&Automatic coding&	M4			
		\\
		
		Classroom observation by observers&	Manual coding&	M3
		\\
		
		\multirow[t]{2}{5.5cm}{Automatic behavior recording}
		&	Manual coding&	A6, A15, V3, V12\\
		&Automatic coding&	A8, A9, A10, A12, A20, V6, V14				
		\\
		
		\multirow[t]{2}{5.5cm}{Mixed recording methods combining manual and automatic methods}
		&Manual coding	&A21, V7\\
		&Automatic coding&	V8, M2			
		\\
		\hline
	\end{tabularx}
\end{table*}

The learning system can be finally represented in front of learners and instructors by considering the representation in the learning specification stage, concerning the factors such as the HCI methods, as well as the behavior recoding and recognition methods (see 5.1-5.2 in Table \ref{Table2}).

According to Table \ref{TableA1}, the typical human-computer interaction methods established in the immersive learning systems were strongly associated with the hardware used and the learning missions designed. In the AR condition, “AR image recognition, and interaction with virtual models through mobile device touch screen” was the most common interaction method for learners to be immersed into the AR environments since most of the AR studies used mobile devices as user interfaces, as illustrated in Fig \ref{Fig5}. In the VR condition, “head and handheld controller movement detection by motion and infrared sensors, and interaction with virtual models using handheld controllers.” for learners using HMDs, and “interaction with virtual models using computer keyboard/mouse and computer screen” for learners using PCs were most popular HCI methods. In the MR condition, “head motion detection by the sensors in OST-HMD and interaction with virtual models using hand gesture manipulation” was utilized for learners using OST-HMD as the hardware medium with MR learning space. To this end, the learners could explore and observe the superimposed virtual information onto the real environments through optical see-through displays and simultaneously interact with the virtual models using gestures (hands-free) \citep{Dosoftei2023,Prilla2019}.

In relation to the behavior sequence recording and recognition methods in the immersive learning environments, the details and the corresponding paper code are tabulated in Table \ref{Table5}. The videotaping combined with manual coding methods (n=17) was the most common in dealing with learners’ behavior sequence issues. The automatic recording and coding methods (n=7) were the second most applied, which is expected to eliminate the outside interference to learners to perform learning activities, decrease human errors by the software tools, and reduce human labor. Apart from videotaping, several other manual behavior recording methods emerged in the review process, including audiotaping combining transcribed verbatim \citep{Lin2013,Wang2014,Wu2019}, which is usually used together with videotaping simultaneously to record learners’ interaction sequences, and classroom observation \citep{Wan2021}. Four studies used mixed recording methods combining manual and automatic methods (n=4) to record the behavioral data of learners for precision and cross-validation.

Overall, the proportion of identified articles using automatic methods to handle the behavior recording (37.5\%) and recognition (32.5\%) was smaller than the proportion of articles using manual methods, which posed a challenge for future researchers to investigate more effective automatic methods to record and recognize interaction behaviors in immersive learning environments.

\subsection{RQ6. What behavioral patterns were constructed in  immersive learning environments?}

The learning evaluation acts as the outcome stage \citep{Garris2002} in BAILF. To thoroughly investigate the learning outcomes of behavioral analysis in immersive learning environments, two key concept matrix facets need to be carefully scrutinized from the identified articles: behavioral analysis methods, and behavioral pattern outcomes (see 6.1-6.2 in Table \ref{Table2}).

As one of the most important reviewed focus in this work, behavioral patterns have been utilized pervasively in extracting specific behavioral patterns by examining the short-term temporal heterogeneity of learning activities \citep{Lamsa2021}. Researchers in the included papers constructed variant behavioral patterns (see Table \ref{TableA1}) by considering the following factors: the behavioral analysis techniques used in analyzing the patterns, the research theme, and the technological features used in learning context designing.
\subsubsection{Techniques used in constructing behavioral patterns}
Regarding the behavioral analysis techniques, various behavioral pattern structures and visualizations were generated by adopting different behavior sequence handling methods. Using behavior frequency analysis (n=5), the basic model of interaction sequences can be represented by the distribution, percentage, and frequency of coded behaviors, and relationships between the basic models can be tested by various data analysis methods, such as t-test \citep{Wu2019}, and correlation analyses \citep{Yilmaz2016}. QCA (n=6) modeled the behavioral content by providing the distribution, frequency, and count information of coded behaviors and was usually used together with other behavioral analysis techniques (e.g., cluster analysis \citep{Cheng2014}, and LSA \citep{Chiang2014,Wang2022}) to complement reliable measurement data. As the most prevalent behavioral analysis method used in the included papers, LSA (n=29) uncovered the chronological relationship between those event sequences that occurred most frequently, and visualized the behavioral patterns using a behavioral transition diagram. For example, \cite{Chen2021b} conducted an LSA and constructed the behavioral transition diagram of teacher-student interactions to reveal behavioral differences in the progressive question prompt-based peer-tutoring approach in VR contexts. SNA (n=2) constructed the social relationships between the group of participants to reveal which member was situated in the central position of the network at different learning activities and visually presented the relationships in the form of social learning network. For instance, \cite{Lorenzo2012} conducted an SNA in an online MR learning platform and found clear social connection between the elected tutor with other learners through the learning network. The network used Freeman’s betweenness to measure the possibility of regulating information flow within the network, and constructed distributed-coordinated learning structure to inform that members were likely to produce interactive impact on each other in the MR learning context. Cluster analysis (n=5) was also used to classify learners’ behavior sequences according to the learning or behavioral characteristic defined in the coding schemes. For example, \cite{Cheng2014} used cluster analysis to identified behavioral features of specific groups of children and parents in the AR picture book reading.
\subsubsection{Behavioral patterns reflecting research themes}
In terms of the research theme, the behavioral patterns of immersive learning situations have been investigated in a wide range of educational topics in the included literature, such as problem-solving collaborative learning \citep{Cheng2019b}, inquiry-based discussion \citep{Chiang2014}, dual-scaffolding learning \citep{Hou2021}, learn by doing \citep{Chen2020}, and co-creation \citep{Wang2022b}.
\subsubsection{Behavioral patterns reflecting immersive features}
In terms of the technological features, behavioral patterns of immersive technologies tended to reflect the interaction between learners and immersive hardware user interfaces. For example, in the AR learning condition, \cite{Hou2021} designed a behavioral analysis that involved the learners' interaction with hardware user interfaces in an AR board game and found that the dual-scaffolding mechanism represented by learners' interactions with makers through mobile devices could facilitate cognitive learning and peer interaction. In the VR condition, \cite{Yang2019} explored the creative process by building the relationships between the brainwave status and learning behavior in a VR environment, in which participants’ painting behaviors using handheld controllers in the immersive environments were recorded by video and EEG and analyzed using LSA. In the MR condition, \cite{Wu2020} analyzed the learners’ behaviors using HoloLens in construction education to obtain learning productivity information.

\begin{table*}[!h]
	\renewcommand{\arraystretch}{1.5} 
	\caption{The challenges in immersive technologies use within the behavioral analysis.}
	\label{Table6}
	\centering
	\scriptsize
	\begin{tabularx}{\textwidth}{ p{3.5cm} p{9cm}X }
		\hline
		\textbf{Challenge categories}               &  \textbf{Challenge description}   &        \textbf{Paper code}\\
		\hline
		
		\multirow[t]{7}{3.5cm}{Technology-related challenges}
		&AR software requires excessive effort in designing compatible educational applications&	A1\\
		&Low stability and correctness in AR marker recognition	&A2, A9\\
		&Highly immersive VR can distract students' attention or Hawthorne effect&	V1, V14\\
		&Simulator sickness	&V1\\
		&Huge tools and contexts differences between VR learning and practical applications&	V2\\
		&VR learning system stability problem	&V5\\
		&The novelty effect of emerging VR technologies	&V13, V14		
		\\
		
		\multirow[t]{8}{3.5cm}{Implementation-related challenges}
		&Small sample sizes&	A4, A7, A15, A21, V5, V6, V9, M4, M5\\
		&Research time restriction&	A5, A12, V6, V13, V14\\
		&Absence of control group&	A7, A11, A15, A21\\
		&Equipment's quantitative restrictions&	V4, V5\\
		&Knowledge diffusion between groups	&V4\\
		&Unequal gender ratio&	V4, V11\\
		&Unfriendly MR manipulation and context settings for young children&	M1\\
		&Low participation or response rate.	&M3, M5
		\\
		
		\multirow[t]{4}{3.5cm}{Analysis-related challenges}
		&Insufficient interaction of learners with physical learning environment using AR system	&A19\\
		&Special behavior sequences recording restriction&	V4, M4\\
		&Insufficient recording due to a limited number of equipment or observations&	V9, M3\\
		&Lack of observation among group members using other behavioral analysis methods&	V11		
		\\
		\hline
	\end{tabularx}
\end{table*}

\subsection{RQ7. What are the challenges in analyzing learners' behavior in immersive learning environments?}

Notably, the BAILF should be regarded as an iterative model to inform and reflect the work-in-progress research of behavioral analysis in immersive learning environments. The learning iteration reflects the planning practices of practitioners and promotes learning effectiveness by structuring systematic judgments \citep{DeFreitas2006}. However, only a few studies reported a clear iteration process in designing and implementing behavioral analysis in immersive learning environments. \cite{Lin2019} customized new coding schemes based on the cluster analysis of first-round behavioral data processing to get the specific and accurate coding scheme suitable for LSA in their investigation. \cite{Sarkar2020} conducted the pilot study before the main studies to validate data instruments, timing, and execution. Two additional studies were carried out as the learning iterations, where the first study verified that learners preferred to perform AR activities in dyads, and the LSA was conducted in the second study with students in dyads. The study with multiple iteration design showed promise for the effectiveness and usability of system construction and an in-depth understanding of the behavioral patterns. 

Additionally, difficulties encountered by researchers and the challenges imposed by immersive techniques in educational settings were also synthesized to provide prospects for researchers planning to investigate learning behavioral patterns in immersive environments in future research(see Table \ref{Table6}). 

\subsubsection{Technology-related challenges}
The technology-related challenges referred to the challenges that researchers encountered when involving immersive technologies in educational settings. For AR condition, the reported challenges were that “the AR marker recognition showed low stability and correctness” \citep{Chang2014,Hwang2018} and “AR software requires excessive effort in designing compatible educational applications” \citep{Cai2021}. For VR condition, the reported challenges showed that the VR technology might bring prohibitive adverse effects on learning, including the Hawthorne effect \citep{Yang2018}, simulator sickness \citep{Chang2020}, novelty effect of emerging technologies \citep{Yang2019}. 

\subsubsection{Implementation-related challenges} 
Implementation-related challenges referred to the challenges that researchers encountered when implementing the learning activities in practical contexts, and the most common challenges were small sample sizes (n=9), research time restriction (n=5), and absence of the control group (n=4). 

\subsubsection{Analysis-related challenges}
The analysis-related challenges referred to those encountered by researchers in recording and analyzing learners’ behavior sequences, as the researcher had difficulties gathering requisite behavioral data. The function of recording interaction sequences of learners in a physical learning environment is hard to integrate into the holistic AR system \citep{Zhang2020}. Similarly, some kinds of unique behavior sequences, such as implicit thinking behaviors \citep{Chen2020}, eye movement, and focus of attention \citep{Wu2019}, are also hard to record through traditional videotaping. Insufficient behavior recording due to the limited number of equipment or observations would be the common restriction when conducting practical behavioral analysis practice \citep{Wan2021,Wang2022}.

\section{Discussion}

In this study, 40 papers were retrieved from four databases for further analysis to uncover behavioral analysis's potential and practical implementations in immersive learning environments. The overview of included studies shows that behavioral analysis in immersive learning environments has gained momentum in recent years.  Based on the constructed framework BAILF and the empirical evidence retrieved from literature analysis, significant findings along with addressing research questions, research limitations, and future research agenda, are discussed in this section.

\subsection{ Future research agenda}

The results by describing literature evidence from the four perspectives are discussed to provide future research agenda in a systematic and integrated view.

\subsubsection{Focusing on meeting learning requirements}

The review evidence about learning requirements from learning stages, cognitive learning objectives/outcomes, and learning activities suggests that some significant research gaps are to be bridged. First, it is worth noting that all of the literature factors about learning requirements were manually synthesized and consolidated by the authors of this review, indicating that few concluded studies had proposed clear pedagogical requirements in the article content. It is necessary to provide salient requirements by defining what stage the learner is at, envisaging what cognitive objectives have to be achieved, and designing an appropriate set of learning mini-activities, reflecting the pedagogical “affordances” before implementing the systems into the specification \citep{Fowler2015}. In addition, the design of mini-activities should take into account several factors systematically, including technical affordances, data processing methods, and intended outcomes, especially the behavioral measures in this review. As shown in Table \ref{TableA1}, representative mini-activities were summarized from the article content of each paper. In these cases, the list of mini-activities usually had narrative features to illustrate the technical affordances of immersive technologies. However, the majority of articles were absent in establishing a close relationship between mini-activities and behavioral measures, resulting in the incompleteness of the whole learning practice. For example, the coding schemes developed by authors or revised based on previous literature are commonly held irrespective of the representative mini-activities designed in the learning systems.

\subsubsection{Elaborating on learning specification}

This review makes efforts to explore the state-of-art practical implementations of different types of learning specifications from identified literature, looking at how to design specific learning activities to analyze learner behaviors within the immersive environments by integrating 4DF into the BAILF to classify review content into four broad categories: learner specifics, pedagogic considerations, context dimension, and representation dimension.

Regarding the learner specifics, the distribution of learner types and application domains revealed that the demand for analyzing how learners interact in immersive learning environments has arisen in a wide range of learner ages and educational domains. Furthermore, the analysis of specific behavioral patterns of learners with special needs in immersive learning environments should be further investigated. The apparently small quantity of studies (only 3) and small sample size of learners (around or less-than 10) in each study remains the standard limitation that needs to be addressed in future research \citep{Parsons2016}.

In terms of pedagogic considerations, the review evidence on the theoretical foundation of each literature was collected from two aspects, including instructional design methods (strategies and techniques), and behavioral coding schemes. Some significant research gaps in pedagogic considerations were discovered in the review process. First, the behavioral analysis and the referenced learning theories were often disconnected, which would prevent the reproducible feature and generalizability of that research \citep{Radianti2020}. For instance, the authors introduced some learning theories to guide the development of immersive learning applications, but the specific pedagogical theory was not reflected in the constructed behavioral patterns. Furthermore, as \cite{Mystakidis2022} suggested, instructional strategies and techniques should be carefully designed to satisfy learning requirements and learner characteristics at different levels. For example, the presentation strategy and observation technique are appropriate options in those learning systems designed for passive conceptual learning. When learners have more freedom to interact actively in the immersive learning context, authors should incorporate learner-center instructional design methods, such as collaborative strategy and game technique. Finally, as shown in Table \ref{TableA1}, the coding scheme design should consider the mini-activities and HCI in the immersive learning system to demonstrate the technical affordance of immersive technologies.

When it comes to context dimension, despite the flexibility and popularity of mobile devices, the AR-based learning system hardware was too singular. The potential behavioral impact of other hardware platforms, such as HMDs and CAVEs, can be explored. Although the number of MR applications was small, there were still ambiguity and a non-homogeneous understanding of the hardware platform that can clearly be considered as “mixed reality” or XR.

Looking at the representation dimension, many papers claim to automatically record and recognize learners’ behavior sequences to minimize the interference of interactivities between learners and virtual worlds. As shown in Table \ref{TableA1}, when planning and envisioning the behavioral patterns, authors should consider the concrete HCI methods that bridge the learner’s physical world with the virtual world of learning space.

\subsubsection{Revealing more profound pedagogical implication through learning evaluation}

This review placed the main focus on deriving the practical application of analyzing how learners interact in the immersive learning context. A few studies evaluated different learning outcomes, such as affective and cognitive outcomes. However, the extra learning evaluation and the constructed behavioral patterns were commonly isolated, lacking a deeper interpretation of the relation between the evaluated factors and behavioral patterns. Moreover, because different behavioral analysis methods can uncover various aspects of learners’ behavioral characteristics, future research needs to combine diverse analysis methods to mine deeper behavioral information. In this review, only the behavioral analysis techniques dealing with behavior sequences assuming short-term temporal homogeneity \citep{Lamsa2021} were included. Other temporal analysis methods that consider the long-term temporal heterogeneity, such as statistical discourse analysis and epistemic network analysis, may also reveal more characteristics of learners’ behavioral patterns in immersive learning environments but have not been thoroughly studied.

\subsubsection{Refining instructional implementation through learning iteration}

In the review process of the BAILF iteration stage, the actual iteration design and implementation examples point out some meaningful indications, which can be used as a reference for the follow-up research:

\begin{itemize}
	
	\item[-] Coding schemes can be designed in an iterative process to better match the coded behavior with the intended learning outcomes of behavioral aspects.
	
	\item[-] Pilot studies and multi-round study designs can be considered to allow instructors to acquire prior information about learners, including preferences for immersive technologies, particular behavioral habits, or design defects, making sufficient preparation for the final major study.
	
	\item[-] The results of various behavioral analysis methods in the same behavioral sequence data can be compared to deeply understand the behavioral differences in the immersive learning environment from different perspectives.
	
\end{itemize}

Additionally, the review results also retrieve some significant shortcomings or difficulties which, however, pave a pathway to potential hazards that require particular attention from scholars, curriculum designers, and software developers in their behavioral analysis in immersive learning environments:

\begin{itemize}
	
	\item[-] More studies dealing with technology-related challenges are needed to enhance the stability and usability of immersive systems. The adverse effects of the immersion learning system on learners should be examined, such as the Hawthorne effect and simulator disease.
	
	\item[-] Regarding implementation-related challenges, small sample sizes, research time restriction, and the absence of a control group are the most severe challenges in the practical exercises and can negatively affect the evaluation of this technology.
	
	\item[-] Regarding analysis-related challenges, advanced instruments, and adequate equipment are needed to record requisite behavioral data for subsequent analysis.
	
\end{itemize}

\subsection{Limitations}

The present systematic review also suffers from some limitations. First, due to the nature of the paper filtering process, some critical publications may be missed based on the inclusion and exclusion criteria. For example, considering that the identified articles were included selectively by focusing on international journals and conference proceedings, it is conceivable that some impressive research from book sections, reports, forums, or working papers could also provide additional insights. Second, though behavioral analysis in the educational environment has been studied for a long time, the study of learners' behavioral patterns in the immersive learning environment is a relatively new academic focus. Since \cite{Cheng2013} advocated adopting mixed methods in analyzing students’ behavioral patterns, relevant papers have only been published in the last decade, resulting in a relatively small review sample size in this study. 

\section{Conclusion}
Considering the fact that little review study has been carried out to systematically synthesize and present the current evidence of behavioral analysis in immersive learning environments. Meanwhile, the conceptual framework has seldom been set up to consolidate knowledge regarding factors emerging from identified articles into a theoretical concept matrix that scholars can analyze, validate, and reuse.
The purpose of this study was twofold. First, considering the deficiency of significant overarching research frameworks in the educational behavioral analysis involving the use of immersive technologies, this study developed a conceptual framework that amalgamates several existing pedagogical models, including the four-dimensional framework, design for learning framework, and IPO framework \citep{DeFreitas2010,Fowler2015,Garris2002}, to prescribe an outlook for future research in practice. 

Second, a systematic review was conducted, on the basis of the proposed theoretical framework, focusing on the research with the intended learning outcomes of behavioral analysis in immersive learning environments. Immersive technologies, learning requirements, learning specifications, evaluation elements, and revealed iterations extracted in the recently published literature on immersive technologies to analyze learners’ behavioral patterns were examined. Possible research hurdles, essential research gaps, and future implications were also examined. It is noteworthy that, throughout this systematic review, in many practical pedagogical applications, it has been found that the technical affordances of immersive technologies and the pedagogical affordances of behavioral analysis were normally isolated and not well integrated into the learning experience. Hence, given the promising future of more widespread usage of immersive technology, further empirical investigations are required to theorize the behavioral impact of learners in immersive learning environments by considering the research agenda proposed in this study. This literature review is anticipated to open up new research avenues and clear a route for scholars and investigators who want to comprehend the present status of behavioral analysis in immersive learning environments and create customized immersive applications based on BAILF for future investigation.



\newpage
\onecolumn

{
	\scriptsize
	\begin{landscape}			
		\begin{longtable}{p{.08\linewidth}  p{0.02\linewidth} p{.25\linewidth}  p{.13\linewidth}p{.15\linewidth}  p{.3\linewidth}} 
			\captionsetup{width=\linewidth}
			\caption{Coded papers and the corresponding Mini-activities, behavior coding schemes, HCI in immersive learning system and behavioral patterns immerged form each paper.}
			\label{TableA1}
			\\ 
			
			\hline
			\textbf{Primary author} 	&  \textbf{Paper Code} & \textbf{Mini-activities} & \textbf{Coding schemes} & \textbf{HCI in immersive learning system} & \textbf{Behavioral patterns} \\ 
			\hline
			\endfirsthead
			
			\multicolumn{6}{r}{\emph{Continued}}\\
			\hline
			\textbf{Primary author} 	&  \textbf{Paper Code} & \textbf{Mini-activities} & \textbf{Coding schemes} & \textbf{HCI in immersive learning system} & \textbf{Behavioral patterns}	\\  
			\hline
			\endhead
			
			\hline
			\multicolumn{6}{r}{\emph{Continued on next page}}\\
			\endfoot
			
			\hline
			\endlastfoot
			
			\cite{Cai2021}& A1	& 
			1. Creating environments, presenting phenomenon;
			
			2. investigation problems, synthesizing phenomenon;
			
			3. sharing explorations, reflecting conclusions.
			&Coding Scheme based on Flanders Interaction Analysis System (iFIAS)&AR image recognition, and interaction with virtual models using virtual buttons.&The behavioral transition diagram of teacher-student interactions revealed that (1) the complexity and flexibility of AR systems would influence learners’ efforts on operation or exploration;(2) learners in the AR system showed enthusiasm.
			\\
			
			\cite{Chang2014} & A2 & 
			1. Descripting information, interpreting knowledge;
			
			2. making judgments; analyzing details;
			
			3. peer discussion.
			&Coding Scheme based on previous literature&AR image recognition, interaction with virtual models through mobile device touch screen.&The behavioral transition diagram of visitors' interactions revealed different behavioral patterns in the nonguided, audio-guided, and AR-guided visitors.
			\\
			
			\cite{Cheng2014} &A3&
			1. Narrating content, reading content;
			
			2. making comments, promoting hints;
			
			3. evaluating comments, expanding responses.
			&Coding Scheme based on previous literature	&AR image recognition, interaction with virtual models through mobile device touch screen.	&Distribution of the coded behaviors through QCA revealed the most frequent behavior; different child-parent behavioral patterns in reading the AR picture book were constructed using cluster analysis.
			\\
			
			\cite{Cheng2016}&A4&
			1. Narrating content, reading content;
			
			2. making comments, promoting hints;
			
			3. evaluating comments, expanding responses.	
			&Coding Scheme based on previous literature	&AR image recognition, interaction with virtual models through mobile device touch screen. &The behavioral transition diagram of child-parent interactions in reading the AR picture book revealed detailed patterns of four cluster groups categorized in A3.
			\\
			
			\cite{Cheng2019b}&
			A5&	1. Answering questions, observing models;
			
			2. problem-solving session.		    
			&Coding Scheme based on Interaction Analysis Model (IAM)	&AR image recognition, interaction with virtual models through mobile device touch screen.	&The behavioral transition diagram of student interactions revealed the patterns at different phases of interaction according to IAM in a problem-solving collaborative learning environment.
			\\
			\cite{Chiang2014}&
			A6&	1. Observing phenomena, recording information;
			
			2. investigation problems, sharing materials;
			
			3. peer discussion, knowledge reflection.&	Coding Scheme based on IAM and its successor.&	Immersive context navigation through sensors embedded in the mobile device, interaction with virtual models through mobile device touch screen.	&The frequency of phase transitions was calculated using QCA; the behavioral transition diagram of student interactions revealed the patterns at different phases of interaction according to IAM in inquiry-based discussion in AR environments.
			\\
			
			\cite{Hou2021}&
			A7&	1. Reading information, observing phenomena;
			
			2. analyzing problems, role playing;
			
			3. collaborative problem-solving, peer discussions.	&Coding Scheme based on previous literature	&AR image recognition, interaction with virtual models through mobile device touch screen.	&The behavioral transition diagram of student interactions revealed behavioral patterns and the differences between high- and low- collective flow groups in the dual-scaffolding AR board game process.
			\\
			
			\cite{Huang2018}&
			A8&	1. Observing models, receiving information.	&Coding Scheme developed by the authors	&AR target recognition, interaction with virtual models through mobile device touch screen.	&The behavioral transition diagram of student interactions revealed gender differences in the AR learning procedure.			
			\\
			
			\cite{Hwang2018}&
			A9	&1. Gathering resources, receiving information;
			
			2. making judgments, investigating problems.	&Coding Scheme based on previous literature	&AR image recognition, interaction with virtual models through mobile device touch screen.	&The behavioral transition diagram of student interactions revealed behavior differences under the active learning-promoting mechanism in the outdoor mobile AR learning context.			
			\\
			
			\cite{Ibanez2016}&
			A10	&1. Observing simulation, discovering facts;
			
			2. designing experiments, drawing conclusions.&Coding Scheme developed by the authors	&AR image recognition, interaction with virtual models through mobile device touch screen.	&The behavioral transition diagram of student interactions visualized behavioral pattern differences when using different scaffolding services.			
			\\
			
			\cite{Lin2013}&
			A11	&1. Receiving information, observing simulation;
			2. constructing relations, applying knowledge.	&Coding Scheme based on previous literature	&AR marker recognition, interaction with virtual models through mobile device touch screen.	&The behavioral transition diagram of student interactions revealed the knowledge construction process of AR dyad learners.			
			\\
			
			\cite{Lin2022}&
			A12	&1. Understanding concepts, observing models;
			
			2. making comparisons, problem-solving;
			
			3. group discussions, reflective thinking.	&Coding Scheme based on previous literature	&AR image recognition, interaction with virtual models through mobile device touch screen.	&The behavioral transition diagram of student interactions revealed behavioral patterns with regard to the contextualized reflective mechanism			
			\\
			
			\cite{Lin2019}&
			A13&	1. Constructing environments, understanding concepts;
			
			2. revealing misconceptions, applying knowledge;
			
			3. group cooperation, reflective thinking.	&Coding Scheme based on previous literature	&AR image recognition, interaction with virtual models through mobile device touch screen.	&The behavioral transition diagram of student interactions revealed behavioral patterns of AR-based science inquiry learning.
			\\
			
			\cite{Matcha2013}&
			A14	&1. Constructing environments, observing models.	&Coding Scheme developed by the authors&	Interaction with virtual objects through manipulating makers.	&The frequency of the interaction and communication, including turn-taking, pointing, parallel and collaborative communication, and gaze, were counted using behavior frequency analysis.
			\\
			
			\cite{Sarkar2020}&
			A15	&1. Presenting visualization, explaining concepts;
			
			2. analyzing relationships, making deductions;
			
			3. abstract deduction, peer discussion.&	Coding Scheme developed by the authors	&Immersive context navigation through sensors embedded in the mobile device, interaction with virtual models through mobile device touch screen.	&The behavioral transition diagram of student interactions revealed learners' behavioral patterns in the AR collaboration learning process by dyads.
			\\
			
			\cite{Wang2014}&
			A16	&1. Receiving information, observing simulation;
			
			2. evaluating facts, applying knowledge.	&Coding Scheme based on previous literature	&AR marker recognition, interaction with virtual models through mobile device touch screen.	&The distribution and frequency of the coded behaviors were obtained through QCA; the behavioral transition diagram of student interactions revealed behavioral differences in collaborative inquiry learning in AR-based and traditional simulation groups.
			\\
			
			\cite{Wang2012}&
			A17	&1. Receiving information, observing simulation;
			
			2. evaluating facts, applying knowledge.	&Coding Scheme based on previous literature	&AR marker recognition, interaction with virtual models through mobile device touch screen.	&The distribution and frequency of the coded behaviors were obtained through QCA; the behavioral transition diagram of student interactions revealed behavioral differences in collaborative inquiry learning in AR-based and traditional simulation groups.
			\\
			
			\cite{Yilmaz2016}&
			A18	&1. Receiving information, discovering facts.	&Coding Scheme based on previous literature	&AR image recognition, interaction with virtual models through mobile device touch screen.	&The frequency of children's interaction and cognitive attainment were obtained using behavior frequency analysis during AR educational magic toy activities.
			\\
			
			\cite{Zhang2020}&
			A19	&1. Receiving information, observing plants;
			
			2. answering questions, making comparisons;
			
			3. peer discussion.	&Coding Scheme developed by the authors&	AR target recognition, interaction with virtual models through mobile device touch screen.	&The behavioral transition diagram of student interactions with physical plants, mobile devices, and other people revealed behavioral differences between the control and experimental groups and the three learning cycle stages.
			\\
			
			\cite{Zhang2016}&
			A20	&1. Entering the system, receiving information.&	Coding Scheme developed by the authors	&AR image recognition, interaction with virtual models through mobile device touch screen.&	The behavioral transition diagram of student interactions revealed behavioral differences of learners with different learning styles, including Visual, Auditory, and Kinesthetic, in the AR-assisted learning activities.
			\\
			
			\cite{Zhang2021}&
			A21	&1. Preparing lessons, receiving information;
			
			2. constructing models, testing knowledge;
			
			3. detailed discussion, collaborative learning.	&Coding Scheme based on IAM	&AR image recognition, interaction with virtual models through mobile device touch screen.	&The visual representation of the learning network was obtained using SNA to reveal which member is situated in the central position of the network at different phases. The behavioral transition diagram of teacher-student interactions revealed the patterns in different phases of interaction according to IAM;
			\\
			
			\cite{Chang2020}&
			V1&	1. Observing models, demonstration knowledge;
			
			2. design generation, creating prototype;
			
			3. sharing experience, peer discussion.&	Coding Scheme developed by the authors	&Head motion detection by sensors in the VR HMD; for zSpace, the stylus pen, and eyewear were traced using depth image detectors, interaction with virtual models using the stylus pen.&	The behavioral transition diagram of student interactions revealed the knowledge construction order in creative VR experiential learning.
			\\
			
			\cite{Chang2021}&
			V2	&1. Mechanism explanation, skill explanation;
			
			2. practicing operation, testing skills;
			
			3. skill assessment.	&Coding Scheme developed by the authors&	Head and handheld controller movement detection by motion and infrared sensors, and interaction with virtual models using handheld controllers.	&The behavioral transition diagram of student interactions revealed the behavioral differences between the experiment and control groups
			\\
			
			\cite{Chen2021b}&
			V3	&1. Observing models, demonstration knowledge;
			
			2. synthesizing findings, answering problems.	&Coding Scheme based on previous literature	&Head motion detection by sensors in the VR HMD.	&The behavioral transition diagram of teacher-student interactions revealed behavioral differences in the progressive question prompt-based peer-tutoring approach in VR contexts.
			\\
			
			\cite{Chen2020}&
			V4	&1. Observation models;
			
			2. exploring topics, explaining understandings;
			
			3. group discussion; performance evaluation.	&Coding Scheme based on previous literature&	In zSpace, the stylus pen and eyewear were traced using depth image detectors, interaction with virtual models using the stylus pen.&	The behavioral transition diagram of student interactions revealed behavioral differences in the hands-on activity combining VR, the 6E model, and STEM education.
			\\
			
			\cite{Cheng2019}&
			V5	&1. Topic instruction, observing content;
			
			2. searching for information, responding to questions.	&Coding Scheme developed by the authors	&Head motion detection by sensors in the VR HMD.	&The count and distribution of the coded behaviors were obtained using QCA; the behavioral transition diagram of teacher-student interactions revealed behavioral differences in different periods over time during virtual field trips.
			\\
			
			\cite{Cheng2015}&
			V6&	1. Understanding knowledge, establishing relationships;
			
			2. role playing, investigating problems.&	Not specified&	Interaction with virtual models using computer keyboard/mouse and computer screen.&	Three core clusters were obtained using cluster analysis according to learners' in-game character use, gaming performance, and concept learning outcomes in VLEs.
			\\
			
			\cite{Ke2018}&
			V7&	1. Responding to instructions, discovering facts;
			
			2. puzzle solving, role-playing;
			
			3. self-identity expression, interpersonal negotiation.	&Coding Scheme developed by the authors	&Interaction with virtual models using computer keyboard/mouse and computer screen.	&Three variant game-based learning patterns of high-functioning autistic youth were obtained using cluster analysis in VLEs.
			\\
			
			\cite{Schmidt2012}&
			V8	&1, Modeling environments, introducing information;
			
			2. practicing skills, applying skills.	&Coding Scheme based on previous literature	&Interaction with virtual models using computer keyboard/mouse and computer screen.&	Mean percentages of coded behavior were counted using behavior frequency analysis and visualized using a bar chart to reveal the specific forms of social behavior of high-functioning autistic youth in the VLEs.
			\\
			
			\cite{Wang2022}&
			V9	&1. Identifying concepts, manipulating simulations; 
			
			2. verifying hypotheses; drawing conclusions;
			
			3. guiding members, peer negotiation.&	Coding Scheme based on IAM and its successor.&	Interaction with virtual models through mobile device touch screen.&	The total count and percentage distributions of coded behaviors were obtained using QCA; the behavioral transition diagram of student interactions revealed behavioral differences of four groups in collaborative inquiry learning in VLEs.
			\\
			
			\cite{Wang2021}&
			V10	&1. Constructing environments, visualizing structure.	&Coding Scheme based on IAM	&Interaction with virtual models using computer keyboard/mouse and computer screen.	&The behavioral transition diagram of student interactions revealed the patterns at different phases of interaction according to IAM in the 3D VR co-creation process.
			\\
			
			\cite{Wang2022b}&
			V11	&1. Constructing environments, comparing information;
			
			2. identifying dissonance, testing outcomes;
			
			3. peer negotiation, making agreements.	&Coding Scheme based on IAM	&Interaction with virtual models using computer keyboard/mouse and computer screen.	&The behavioral transition diagram of student interactions revealed the patterns at different phases of interaction according to IAM in the 3D VR and the 2D co-creation process.
			\\
			
			\cite{Wang2018}&
			V12	&1. Identifying emotions, watching videos;
			
			2. role-playing, making decisions;
			
			3. sharing expressions, peer discussion.	&Coding Scheme based on previous literature	&Interaction with virtual models using computer keyboard/mouse and computer screen.	&Three variant game-based learning patterns of high-functioning autistic youth were obtained using cluster analysis in VLEs.
			\\
			
			\cite{Yang2019}&
			V13&	1. Brainstorming models;
			
			2. designing models, demonstrating outcomes.&	Coding Scheme based on previous literature	&Head and handheld controller movement detection by motion and infrared sensors, interaction with virtual models using handheld controllers.	&The behavioral transition diagram of student interactions according to the brainwave values of attention and meditation revealed the behavioral differences between different product creativity groups in immersive VR environments.
			\\
			
			\cite{Yang2018}&
			V14	&1. Entering contexts, brainstorming models;
			
			2. designing models, interacting with models.	&Coding Scheme based on previous literature	&Head and handheld controller movement detection by motion and infrared sensors, interaction with virtual models using handheld controllers.&	The behavioral transition diagram of student interactions according to the brainwave values of attention and meditation revealed the behavioral differences between paper-and-pencil conditions and immersive VR conditions.
			\\
			
			\cite{Hsu2020}&
			M1	&1. Reading information, interpreting facts.	&Coding Scheme based on previous literature&	Location-based AR and Interaction with virtual models through mobile device touch screen.	&The behavioral transition diagram of caregiver-child interactions revealed social interaction patterns in the MR learning environments. 
			\\
			
			\cite{Lorenzo2012}&
			M2&	1. Constructing environments, presenting models;
			
			2. role-playing, asynchronous interactions;
			
			3.  collaborative evaluation, group discussion.	&Not specified&	Interaction with virtual models using computer keyboard/mouse and computer screen.	&The visual representation of the learning network was obtained using SNA to reveal how the online tutor is situated in the central position of the network in the MR learning platform context.
			\\
			
			\cite{Wan2021}&
			M3	&1. Making announcements, showing demonstrations;
			
			2. providing feedback, posing questions;
			
			3. class discussion.	&The modified version of the Laboratory Observation Protocol for Undergraduate STEM (LOPUS)	&Body motion tracking by the motion-sensing input device ($Kinect^{TM}$).	&Graduate teaching assistants’ instructional style in MR learning environments were categorized by cluster analysis
			\\
			
			\cite{Wu2019}&
			M4&	1. Gathering information, interpreting facts;
			2. exploring models, developing values.	&Coding Scheme developed by the authors&	In VR, head and handheld controllers’ movement tracking by motion and infrared sensors; In MR, head motion detection by the sensors in OST-HMD and interaction with virtual models using hand gesture manipulation.	&The distribution of coded behavior was counted and analyzed using a t-test to identify the accessibility issues of student novices and professional experts in both VR- and MR-constructed virtual environments
			\\
			
			\cite{Wu2020}&
			M5	&1. Building structure, gathering information;
			2. exploring models, developing values.&	Coding Scheme developed by the authors	&Head motion detection by the sensors in OST-HMD and interaction with virtual models using hand gesture manipulation.	&The distribution of the corresponding time durations of coded behavior was calculated to identify productivity in MR-constructed virtual environments and paper drawings manners.
			\\
			
		\end{longtable}
	\end{landscape}
}

\newpage
\twocolumn

\bibliographystyle{apalike}


\section{Competing interests}
The authors declare no competing interests.

\section{Ethical approval}
This article does not contain any studies with human participants performed by any of the authors.

\section{Informed consent}
This article does not contain any studies with human participants or animals performed by any of the authors. Informed consent is not applicable.

\section{Data availability}
All data generated or analyzed during this study are included in this article and its supplementary information file.

\section{Figures legends}

\begin{itemize}
	
	\item \textbf{Fig. 1 Behavioral analysis in immersive learning framework (BAILF)}
	
	\item \textbf{Fig. 2 Literature identification process derived from the PRISMA framework}. Using the previously-defined key concept terms, this review yielded 676 results from databases. With the aid of the StArt software, 53 duplicate papers were found and deleted. During the screening phase, a review of the titles, abstracts, and keywords revealed 560 irrelevant articles, and 63 studies that met the inclusion criteria were included for the following selection stage. During the eligibility phase, the entire text of the remaining articles was scanned in detail to check the theoretical contribution in the area of learning implementation, behavior analysis, virtual communities, and the extension of instruction theories. Thus, 24 studies with limited evidence on analyzing learner behavior sequences or without immersive intervention were excluded. Additionally, another five articles were classified as ineligible due to the lack of reliability in their contributions or inconsistent analysis. Then, the backward and forward snowballing method was carried out on Google Scholar to find more relevant literature, including another six papers in this review. In the end, 40 papers were ultimately classified as eligible and included in the final review.
	
	\item \textbf{Fig. 3 Learning stages and cognitive learning outcomes}. In the bottom right bar chart, a majority of researchers have implemented immersive learning for the higher echelons of learning stages, i.e., dialogue (n=20), followed by construction (n=14) as the mediate echelons of learning stages, and conceptualization (n=6) as the lower echelons of learning stages. In the upper-left bar chart, the most common cognitive learning outcome based on Bloom’s revised taxonomy was the ability to create knowledge (n=13). Remembering (n=5), understanding (n=7), applying (n=6), and analyzing (n=5) were the following most common cognitive learning outcomes that participants achieved in the immersive learning environments. Finally, in 4 studies, participants acquired the ability to evaluate the immersive learning activities.
	
	\item \textbf{Fig. 4 Learner specifics}. \textbf{a},  learner types. The study participants were mostly primary school students, with 12 papers accounting for 30\% of the total, and higher education students, with 9 papers accounting for 22.5\% of all papers. Other studies tended to recruit learners of adults (17.5\%, n=7), high school students (12.5\%, n=5), kindergarten children (10\%, n=4), middle school students, teachers (7.5\%, n=3), and students with special needs (7.5\%, n=3). One paper, accounting for 2.5\% of the total, did not specify the learner type in the article content. \textbf{b}, application domain. More than half of the articles chose STEM (62.5\%) as the application domain of their learning systems. The second popular application domain was general knowledge \& skills (27.5\%), where learners can learn basic social or art abilities to deal with daily affairs. The rest of the articles chose to learn about knowledge in the humanities (10\%). Specifically, in the category of STEM, physics (17.5\%), integrated science (17.5\%), and biology (12.5\%) were popular topics in the immersive learning system. “Integrated science” refers to the specific subject that involves more than one scientific discipline in the learning activities. In the humanities category, history (2.5\%), culture (2.5\%), and language learning (5\%) were the common topics. In the general knowledge \& skills category, the literature highlighted three topics of educative applications: art \& design (7.5\%), cognitive \& social skills (12.5\%), and reading (7.5\%).
	
	\item \textbf{Fig. 5 Hardware devices}. As for AR condition, except for one paper that used desktop computing devices (n=1) to construct the learning system, all other papers used mobile devices (n=20) as predominant apparatus. As for VR conditions, the non-immersive VR was equipped with mobile devices (n=1) and desktop computing devices (n=6). Full-immersive VR that adopted HMD devices (n=7) as infrastructure was widespread in the immersive learning system construction. Two studies used zSpace as hardware devices, which is expected to provide learners with a semi-immersive experience in this review. As for MR conditions, four kinds of hardware devices were used to set up MR systems: mobile devices (n=1), desktop computing devices (n=1), OST-HMD (n=2), and projection-based apparatus (n=1).
	
\end{itemize}

\section{Equipment and settings}

\begin{itemize}
	
	\item Fig. 1 is processed using Microsoft Visio.
	
	\item Fig. 2 is processed using Microsoft Visio.
	
	\item Fig. 3 is processed using Microsoft Excel and Visio.
	
	\item Fig. 4 is processed using Microsoft Excel and Visio.
	
	\item Fig. 5 is processed using Microsoft Excel and Visio.
	
\end{itemize}

\end{document}